# Industrial applications of digital rock technology

Carl Fredrik Berg[1*], Olivier Lopez[1] and Håvard Berland[1]

[1] Statoil R&D Center, Arkitekt Ebbels veg 10, Rotvoll, 7053 Trondheim, Norway

[*] Corresponding author's email address: carl.f.berg@ntnu.no. Present address: Department of Geoscience and Petroleum, NTNU, Trondheim, Norway.

## Abstract

This article provides an overview of the current state of digital rock technology, with emphasis on industrial applications. We show how imaging and image analysis can be applied for rock typing and modeling of end-point saturations. Different methods to obtain a digital model of the pore space from pore scale images are presented, and the strengths and weaknesses of the different methods are discussed. We also show how imaging bridges the different subjects of geology, petrophysics and reservoir simulations, by being a common denominator for results in all these subjects. Network modeling is compared to direct simulations on grid models, and their respective strengths are discussed. Finally we present an example of digital rock technology applied to a sandstone oil reservoir. Results from digital rock modeling are compared to results from traditional laboratory experiments. We highlight the mutual benefits from conducting both traditional experiments and digital rock modeling.

## 1  Contents

## 2 Introduction

Information about subsurface reservoir characteristics can roughly divided into two classes; data obtained from surface surveys such as seismic, electromagnetic, gravimetric etc., and data obtained from wells, such as well logs, transient pressure testing, reservoir performance, cores and drill cuttings. While data obtained from the surface typically have low resolution and large coverage, well data typically has high resolution and low coverage. At the extreme, core material can provide pore scale resolution, the foundation for digital rock physics, as will be discussed in detail in this article.

Cores provides essential data for formation evaluation, with direct information on the presence, distribution and transport of reservoir fluids at a peerless resolution (Andersen et al., 2013). Conventional core analysis (CCA) provides basic properties such as porosity, permeability and grain density, while special core analysis (SCAL) provides more complex reservoir properties such as capillary pressure, relative permeability and electrical properties (McPhee et al., 2015). Digital rock physics is integrated with and provides similar types of data as CCA and SCAL, and consequentially has a similar entry point into conventional subsurface workflows as CCA and SCAL. However, the strong modeling component of digital rock physics brings it closer to reservoir modeling than traditional CCA and SCAL.

Digital rock physics (DRP) is to numerically solve transport equations on digital representations of rock samples. This roughly divides digital rock technology into two parts: imaging and modeling (Andrä et al., 2013a, 2013b; Blunt et al., 2013). The aim of digital rock physics is thus to derive a digital description of a reservoir rock, and then perform numerical simulations of transport processes on this obtained digital structure. It could be argued that digital image analysis is a prerequisite for digital rock physics, whereas digital rock physics only comprises simulation of rock physics. Such a distinction would classify the calculation of porosity as digital image analysis, while obtaining the permeability by solving the Stokes equation on a segmented rock image would classify as digital rock physics. Such a distinction have advantages, however imaging and simulations are so intertwined that this article will classify both as digital rock physics.

Traditional workflows for rock physics are typically based on purely empirical relations from laboratory experiments, e.g. Archie's law (Archie, 1942). Theoretical models based on simplified microstructures and calibrated to laboratory data is also commonly used, e.g. the Kozeny-Carman relationship (Carman, 1937; Kozeny, 1927). There are limitations for the number of different data types that can be measured on a single rock sample with traditional laboratory experiments. In contrast, any effective property that can be calculated with digital rock modeling can be calculated on the same pore structure, i.e. the same sample (Arns et al., 2005). This removes a major uncertainty for cross-property relations: The assumed similarity of sister-plugs. Further, digital rock physics opens up new possibilities to link effective properties to the underlying pore structure, and thereby obtain fundamental cross-property relations (Berg and Held, 2016). Furthermore, pore scale characterization of two-phase flow can lead to fundamental understanding of macroscopic parameters such as relative permeability (Liu et al., 2017). Calculation of transport properties on digital representations of the pore structure are typically fast compared to laboratory experiments. Calculation of different properties can be conducted in parallel, thus enabling a further acceleration compared to laboratory experiments if the same plug is planned to be used serially in several experiments. In summary, digital rock physics has the possibility to reduce uncertainty in formation evaluation by providing fundamental rock physics relations at a reduced time.

Digital rock physics typically require imaging and modeling at high resolution and large coverage (Andrä et al., 2013a, 2013b). Due to limitations on imaging and computational power, the advent of digital rock technology was mostly devoted to basic research on transport and displacement processes on simplified pore structures or on clean well-sorted outcrop rock samples. Improvements in imaging and computational power has enabled application of digital rock physics on ever more complex rock samples. Such improvements has transformed the application of digital rock technology from mostly being devoted to basic research to an integrated element in (oil-)industry workflows for modeling of transport in porous media (Blunt et al., 2013; Fredrich et al., 2014; Lopez et al., 2012, 2010).

The obtainable properties from digital rock physics can be divided into three classes. The first class is mainly dependent on an accurate description of the pore structure. With an accurate three-dimensional (3D) pore structure description one can obtain conventional core analysis single-phase properties such as porosity and permeability (Berg, 2014; Øren et al., 2007; Schwartz et al., 1993). Also the formation factor is mainly dictated by the pore structure, however it should be noted that formation factor could be significantly influenced by clays and surface conductance (Berg, 2012; Johnson et al., 1986; Wildenschild et al., 2000). The second class is multiphase properties with simple fluid-fluid and fluid-rock interactions, such as capillary dominated primary drainage. This includes properties such as primary drainage capillary pressure (including mercury injection) and saturation (electrical resistivity) index (Bekri et al., 2003; C. F. Berg et al., 2016; Gomari et al., 2011). The third class has complex rock-fluid and fluid-fluid interactions, such as relative permeability and imbibition capillary pressure (Øren et al., 1998; Ramstad et al., 2010; Valvatne and Blunt, 2004).

The two first classes of digital rock results are thus mainly influenced by the digital representation of the pore structure, while the third class additionally needs accurate information on fluid-fluid and fluid-rock interactions. These two types of input translates into the two main challenges for digital rock physics.

The first is to obtain 3D information which resolves the pore structure to a sufficient degree, and then to balance the opposing objectives of sufficient resolution with a representative elementary volume (REV) (Al-Raoush and Papadopoulos, 2010). This challenge is strongly dependent on the sample; clean sands with small grain size variation is demonstrated feasible today, while larger grain size variation, cementation, clays and partly dissolved grains still pose significant problems. Advances in imaging has been an integral part of the growth of digital rock modeling, and has to some extent solved the challenge of obtaining pore structure data of sufficient resolution. Larger images translate into computational challenges; modeling and simulations with sufficient resolution at a REV is computational heavy. This challenge is commonly targeted by brute force and parallel computing, while dynamic gridding, network modeling, hybrid solutions and upscaling are other ways to tackle the same issue.

The second main challenge for digital rock physics is how to implement fluid-fluid and fluid-rock interactions. These properties are translated into interfacial tension, wettability and contact angles, thus one need interfacial tension, wettability and a contact angle distribution representative for the reservoir in question (Sorbie and Skauge, 2012). The fluid-fluid property interfacial tension is fairly straightforward to measure and to implement into simulators. For the fluid-rock properties, there is a range of methods to include wettability and contact angle descriptions into pore scale multiphase simulations (Blunt, 1998; Øren et al., 1998; Valvatne and Blunt, 2004). Thus the challenge is not a lack of flexibility when including fluid-rock interaction into the simulations, but how to decide on a representative description. There exist fairly simple measurement methods for fluid-rock properties, such as measuring the contact angle of a droplet on a surface. Unfortunately it is not simple to implement the experimental results into pore scale simulators with a complex surface geometry and a range of mineral surfaces. An even harder problem, which is shared with traditional experiments, is to relate the fluid-rock properties to reservoir properties. Typically this is solved by ageing rock samples in crude oil from the relevant oil reservoir. It is assumed that the rock surface then changes wettability to a state that resembles reservoir conditions. The fluid configuration at such a wettability state can be imaged, both in micro-tomography (micro-CT) and with scanning electron microscopes (SEM). Contact angles can be measured directly by imaging, at least for samples were one of the fluids is strongly wetting (Andrew et al., 2014). Unfortunately, such advanced imaging cannot answer the underlying problem: Is the obtained wettability state representative for the reservoir in question.

As digital rock models are many orders of magnitude smaller than a grid block in a reservoir simulation model, bringing digital rock results into a full field reservoir simulation involves upscaling (Aarnes et al., 2007; Lohne et al., 2006). This is a challenge digital rock modeling shares with experimental results from core samples: Relative to the size of grid blocks in reservoir models the size of core samples and digital rock models are essentially the same. As geological structures might contain laminas smaller than the size of core plugs, flow parameters obtained from core flooding does not always correspond to a REV (Bhattad et al., 2014; Nordahl and Ringrose, 2008). Digital rock modelling can easily accommodate different model sizes, and thereby ensure the flow parameters are representative for a REV (Nordahl et al., 2014; Odsæter et al., 2015).

Digital rock physics has been applied to a variety of sedimentary rocks. Arguably, digital rock physics has been centered on siliciclastic rocks. The advent of the technology focused on outcrop samples (Arns et al., 2001; Bakke and Øren, 1997; Schwartz et al., 1994), while subsequent studies has been extended to reservoir rocks (Lopez et al., 2010; Øren et al., 2007). Carbonates, the second most abundant rock type for oil reservoirs, has also been studies from the onset of digital rock physics (Schwartz et al., 1994), with later uptake for industrial applications (Grader et al., 2010; Kalam et al., 2012; Lopez et al., 2012). It should be noted that large pore size variations in carbonates might require multiscale modeling (Biswal et al., 2009; Lopez et al., 2012). The arrival of the shale hydrocarbon industry has delivered a third class of reservoir rocks investigated by digital rock physics. Recently a majority of cored wells in the US are from shale reservoirs, making shale an important rock type for core analysis (Walls and Armbruster, 2012). The extremely small pore sizes in such samples make them intrinsically hard to resolve and model (Kelly et al., 2016). However, as such samples are equally hard to measure using traditionally laboratory procedures, digital rock physics has been a popular choice for evaluating shale samples (Walls et al., 2012).

This article starts with an overview of different imaging techniques and 3D reconstruction that has been applied in digital rock technology. Section 4 presents methods and results from image analysis, including direct industrial applications of image analysis results. In Section 5 we present different pore scale simulation methods, with a focus on the direct modeling versus network modeling. An application of digital rock technology is presented in Section 6, while we summarize this article in the last section, Section 7.

# 3 Imaging

In digital rock physics workflows, imaging of rock samples typically has a two-fold purpose: The first is to obtain sample description such as grain size distribution, pore size distribution, mineralogy, clay content etc. The other purpose is to obtain 3D structures for computation of effective properties such as permeability and conductivity. Those two imaging purposes cause different constraints, some of which are conflicting.

At the early stage of digital rock physics it was more common to reconstruct 3D structures from 2D electron microscopy images than to use 3D X-ray tomography images directly. This was due to the low quality and low availability of X-ray tomography at the resolution needed for predictive modeling. As X-ray tomography is of better quality and more readily available today, it is more frequently used. However, there are situations where the advantages from starting with 2D images make this method preferable also today.

## 3.1 3D imaging

X-ray computed tomography (CT) reconstructs a 3D image of a sample from a set of X-ray images of the sample taken at different angles (Wildenschild and Sheppard, 2013). CT-imaging has been used in the oil industry for decades (Cromwell et al., 1984). Typical use includes core inspection, localization of where to drill plug samples, inspection of core plugs, imaging of core flooding experiments and rock characterization (Gilliland and Coles, 1990; Honarpour et al., 1985; Hove et al., 1987; Wellington and Vinegar, 1987). For such applications it is common to use a medical CT scanner. In medical CTs the X-ray source and detectors are rotating around the stationary sample.

Calculation of effective rock properties from digitized images of the pore structure requires an image resolution that sufficiently resolves all parts of the structure important for the property in question. Digital rock physics therefore requires a higher image resolution that what is obtained using standard medical CTs. Some of the first high-resolution images of rock samples were acquired using the beam-line at a synchrotron facility as the X-ray source (Flannery et al., 1987). Today, imaging using synchrotrons are mostly conducted when one wants to capture rapid transient processes on the pore scale (Berg et al., 2013).

Laboratory based micro-CT equipment are readily available today, and is the standard equipment for obtaining 3D images of rock samples for digital rock physics (Wildenschild and Sheppard, 2013). Most such laboratory micro-CTs use a stationary X-ray source and detector while the sample is rotated, however micro-CT gantry scanners are also used (Bultreys et al., 2015).

As stated, when imaging for digital rock analysis one needs to resolve the structure important for the property of interest. Exemplary, the main flow paths are of ultimate importance for calculation of flow properties, while electrical properties is dependent on all porosity, also the microporosity to be discussed in Subsection 4.2.2.

Further, the 3D distribution of X-ray attenuation values needs to be discretized into a form that allows for calculation of effective properties. For calculation of flow properties this typically means segmentation into pore and matrix phases, and sometimes also a third micro-porous phase. One common segmentation practice is to use a global threshold value, were the threshold might be chosen to enforce a separately measured porosity value (Iassonov et al., 2009). Often the measured porosity is representing a larger sample than the imaged subvolume, or the sample size is so small that the porosity measurement becomes uncertain. More fundamentally, when a significant amount of the porosity is unresolved, the porosity cannot be used as a guide for global threshold (Leu et al., 2014). Information on the unresolved porosity could be obtained by dry-wet imaging, to be discussed in Subsection 4.2.2. Global thresholds approaches without a prior porosity value, e.g. approaches based on histogram analysis or visual inspection, has been shown to be significantly user dependent (Iassonov et al., 2009). However, they can be improved significantly by application of filters before segmentation (Sheppard et al., 2004). Use of local spatial information, e.g. indicator kriging based segmentation, reduce user dependence (Oh and Lindquist, 1999). Local spatial information has been deemed crucial, however such methods still require significant supervision by the operator (Iassonov et al., 2009).

For a coarse grained outcrop sandstone like Fontainebleau with a typical grain size of $200\mu m$ a resolution of around $5\mu m$ can be sufficient, while more complex pore structures typically needs higher resolution (Shah et al., n.d.).

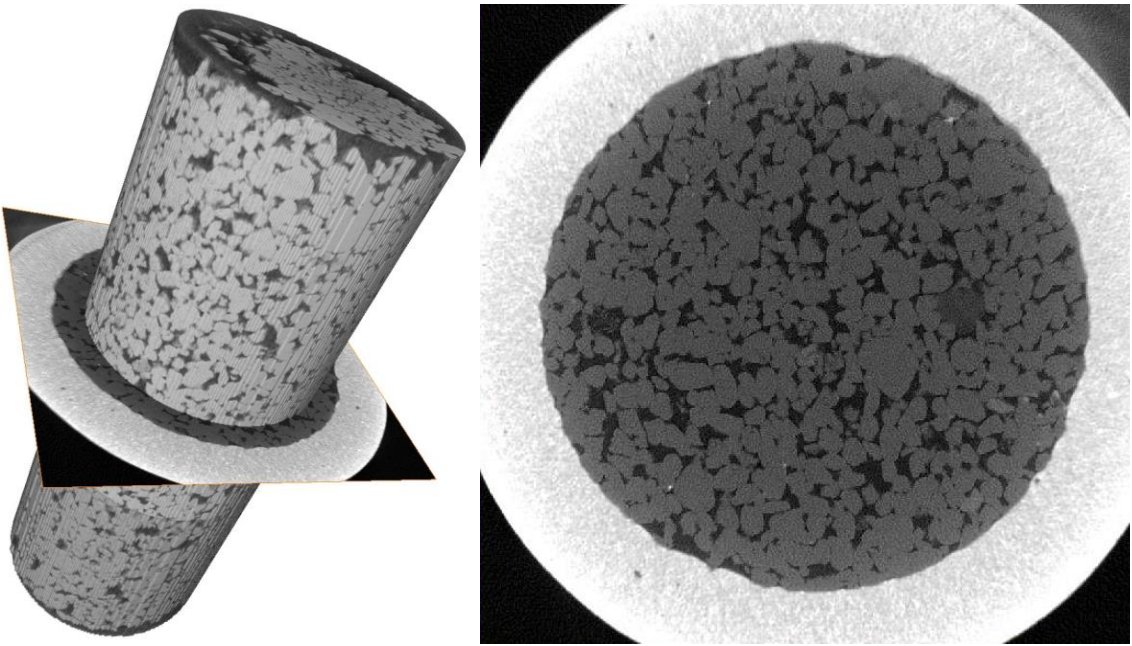

**Figure 1:** A micro-CT image of an outcrop Bentheimer sample. The image has a resolution of 6.7 ***μm***, while the sample has a diameter of approximately 5mm.

A micro-CT image of an outcrop Bentheimer sample is shown in Figure 1, both as a 3D volume and an extracted cross-section. Higher resolution implies smaller sample volume for micro-CT imaging. While the resolution and sample volume of the Bentheimer sample shown here could be considered typical, more complex samples would demand higher resolution at the expense of sample size.

Denser materials are typically higher X-ray attenuating, which is reflected by higher CT-values visualized with lighter colors. On the other hand, materials with low atomic weight gives lower values visualized by darker colors. Thus the air filled pore space appears as black, while the heavier grains appear as gray in Figure 1.

## 3.2 2D imaging

There are several microscopy methods which produces high-resolution images of rock samples. The most common is back-scattered electron (BSE) images from scanning electron microscopes (SEM). BSE is routinely applied to acquire images with resolution in the sub-micron scale, and such BSE images are commonly used in rock characterization. Measuring X-ray diffraction produces additional information on the mineralogy (Minnis, 1984).

Focused ion beam SEM (FIB-SEM) is a newer technology which produces 3D images with the same resolution as SEM (Lemmens et al., 2010). This is a destructive method where the focused ion beam mills thin slices from the sample, while the SEM sequentially acquires images of the newly exposed surface. The resulting 3D image typically covers a volume of around 5 $\mu m^3$, which raises concerns of the representativeness of the sample (Kelly et al., 2016). This method has mostly been applied to shale samples to obtain 3D images of the extremely small pores in such samples (Curtis et al., 2010; Walls and Sinclair, 2011).

A typical workflow for BSE imaging starts by hardening a rock sample with epoxy, where the sample is placed in a vacuum chamber and impregnated with epoxy resin. A thin section is then cut from the epoxy-filled sample. The thickness of the thin section is in the order of 20 $\mu m$, and is fixed to a glass-plate. In principal this is a destructive technique, but in practice it is common to use endcaps from plugs, as such endcaps has limited use.

For optical imaging of thin sections it is common to have a cover glass, while BSE imaging requires an uncovered surface. Thus any cover glass needs to be removed, and the thin section surface should be polished to produce a flat surface. This surface is typically carbon coated before imaging. Backscattered electrons consist of electrons originating from the electron beam, where elements with high atomic number backscatter electrons more strongly than elements with low atomic number. This produces gray scale images, where the epoxy appears darker than the denser grains. As the epoxy has impregnated the pores, the dark areas thus represent the pore space (Dilks and Graham, 1985).

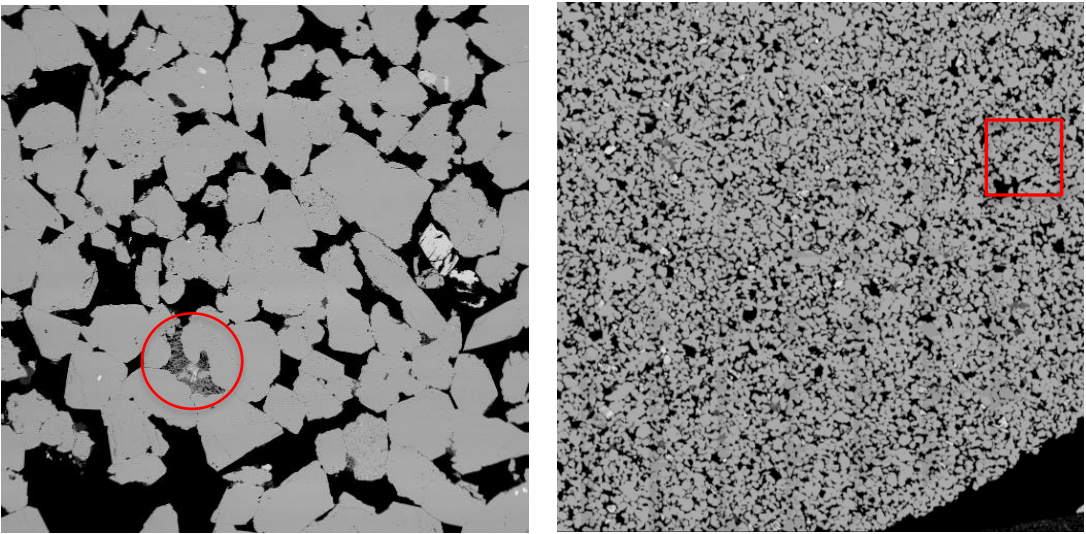

**Figure 2:** BSE images of a thin section of a Bentheimer sample. The sample is circular with a 1.5 inch diameter, with the right image containing part of the circular shape. To the left is an acquired BSE image of the sample surface, containing $512^2$ voxels. Such individual images are stitched together to form an image of the complete sample. There are few artifacts arising from the stitching process, as evident from the image to the right. The circle in the left image indicate occurrence of authigenic clay.

A BSE image of a Bentheimer sample is shown in Figure 2. The image is created with an automated process, where the SEM acquires images of smaller sections of the surface, before these images are stitched together. This automated process creates an image covering the whole sample, and due to low levels of drift the resulting image can directly be employed for quantitative image analysis.

### 3.3   3D reconstruction from 2D images

Digital rock physics calculations employ 3D digital images of the pore structure. As the images presented in Subsection 3.2 merely contains 2D information, it is essential to reconstruct 3D structures from such 2D information. Several reconstruction methods are purely based on statistical information from the 2D section, usually correlation functions (Adler et al., 1992; Okabe and Blunt, 2004). There has recently been several promising efforts using machine learning to reconstruct porous media (Mosser et al., 2017; Srisutthiyakorn, 2016). Reconstruction methods based on statistical information are able to preserve important features of the original samples; in addition they are applicable to a wide range of different porous media.

Rock samples are not random heterogeneous porous media, but have a structure that is the result of a particular rock forming process. Reconstruction methods that includes prior information of the rock sample has successfully been applied to carbonates (Biswal et al., 2009), however it is most commonly applied for clastic sandstones (Øren and Bakke, 2002). Such methods have been applied to generate a 3D laboratory scale model of a Fontainebleau sandstone with a large spread of resolutions (Hilfer et al., 2015; Latief et al., 2010).

Attempting to replicate the rock forming process adds prior information of the rock sample and concurrently connects the resulting pore structure to the geological rock forming process. This meets two objectives; incorporating prior information could improve the 3D reconstruction, while relating pore structure to the rock forming process is valuable for rock characterization.

One process based 3D reconstruction was introduced in (Bakke and Øren, 1997), where three steps are applied to mimic the natural occurring rock forming process: Grain sedimentation, compaction and diagenesis. The grain sedimentation is modelled by creating a sphere pack with a size distribution representative for the grain size distribution. Compaction is modelled by a linear shift of the grains in the vertical direction. The main elements of the diagenesis modeling are to simulate quartz overgrowth, clay and cement precipitation. Quartz might be modelled using a method introduced in (Pilotti, 2000), while clay and cement are usually distributed statistically. Such a process based modeling technique has been shown to be more predictive than typical statistical reconstruction techniques (Hilfer and Manwart, 2001).

# 4 Image analysis

Imaging involves processing to enhance the image quality, e.g. beam hardening correction, removal of ring artifacts, filtering, etc. (Wildenschild and Sheppard, 2013). Such processing is essential to obtain high quality images, however it is considered outside the scope of this paper. In this section we consider image analysis on already processed images.

The input for simulation of transport processes are 3D structures containing few phases, often simply binary images representing only the matrix and pore space (Iassonov et al., 2009). This requires processing of the rock sample images irrespectively of where they are originating from; micro-CT, SEM or other methods. Segmentation into the relevant phases needs to be calibrated, as the segmentation process is non-unique (Andrä et al., 2013a; Leu et al., 2014; Schembre-McCabe et al., n.d.). In addition to an input for simulations, image analysis can provide further information of the samples; e.g. grain- and pore-size distribution, mineralogy, and clay content.

## 4.1 Mineralogy

Mineralogy is one of the key parameters in characterizing rocks. One commonly applied method to quantify the mineralogy is X-ray diffraction measurements of crushed samples (Środoń et al., 2001). This involves crushing the sample to a powder, the method is thus destructive and it gives no information on spatial relationships. Another commonly applied method is optical petrography. Samples can be inspected manually, e.g. by point-counting methods (Walderhaug et al., 2012), or by digital image analysis (Grove and Jerram, 2011).

In both micro-CT and BSE images the gray scales correlates to the density of the material, thus such gray scale images have information on the mineralogy of the sample. Unfortunately, different minerals might have similar gray values, e.g. the BSE values for quartz and plagioclase are similar, hampering mineral identification.

Other methods to obtain more quantitative measurements for mineralogy are available. It is possible to back-calculate density values from dual-energy scans, i.e. a set of two micro-CT scans obtained at different energy levels, however this is more widely used on larger scales such as whole cores (Lopez et al., 2016). For thin sections it is common to apply energy-dispersive X-ray microanalysis (EDS) in combination with SEM to obtain elemental information (Minnis, 1984). Together with a mineral identification system this can give high resolution mineral map of the imaged core sample (Gottlieb et al., 2000; Knackstedt et al., 2010). Mineral quantification can be used in rock typing directly, by separating different rock types based on the amount of different minerals.

## 4.2 Porosity classification

Porosity information is obtained from images by segmentation. For a dry sample in ambient conditions the pore space is filled with air. It is typically straight forward to detect large pore bodies in micro-CT images, as air has low X-ray attenuation compared to minerals such as quartz. Smaller pores, e.g. tight pore throats and pores inside clays, is typically not fully resolved in a micro-CT image. With SEM imaging such pores can be resolved, but higher resolution is conflicting with the objective of obtaining a REV. Exemplary, authigenic clays has pore sizes in the range of 100Å to 1000Å, which are several orders of magnitude smaller than grains (Hurst and Nadeau, 1995). Thus resolving such small pores and at the same time capturing a REV is hardly possible in 2D, while it is clearly outside present capabilities in 3D.

When excluding fractures, sandstone porosity can be classified into three types: Intergranular, dissolution and micro (Pittman, 1979). The integranular porosity is the primary porosity between the grains. Dissolution of soluble rock components, e.g. feldspar grains, creates dissolution porosity. Completely dissolved components can create large vugs, which gives a larger effect on porosity than permeability. Microporosity is commonly defined by size, e.g. pores smaller than $0.5 \mu m$, and occurs e.g. in clays. There might be significant overlap between these porosity definitions. When working with images it is common to classify the porosity after how it is resolved in the image: The resolved porosity is defined as effective porosity, while porosity below the image resolution is called microporosity (Lopez et al., 2013). The separation between effective and micro-porosity is thus dependent on the image resolution and image quality.

The capillary pressure is linked to pore sizes through the Young-Laplace equation:

$$P_c = \frac{2\sigma \cos \theta}{r}, \qquad (1)$$

where $\sigma$ is the interfacial tension, $\theta$ is the contact angle, and $r$ is the radius of the pore. For mercury imbibition into a dry core it is common to assume that the contact angle is constant as mercury is non-wetting for most materials. The interfacial tension is also assumed constant, thus the capillary pressure is inversely proportional to the pore radius from the Young-Laplace equation. One can therefore get information of the pore size distribution by monitoring the capillary pressure during a mercury intrusion experiment (Purcell, 1949). Measurements of high accuracy can actually observe when single pore throats are intruded by the mercury (Yuan and Swanson, 1989).

We can quality control the accuracy of the pore structure for a segmented pore space from a micro-CT image, equivalently the pore space in a reconstructed 3D structure from a 2D image, by comparing a mercury injection simulation to experimental data. Such a comparison is shown in Figure 3. Note that the simulated mercury intrusion saturations have been scaled by the fraction of resolved porosity in the images to the measured porosity on the corresponding plug used in the mercury experiment. The difference between resolved and measured porosity is then attributed to pores smaller than the image resolution. The good match between simulated and experimental data observed in Figure 3 indicates that the image segmentation has captured the pore structure to a high degree of accuracy. This yields a good starting point for further numerical simulations.

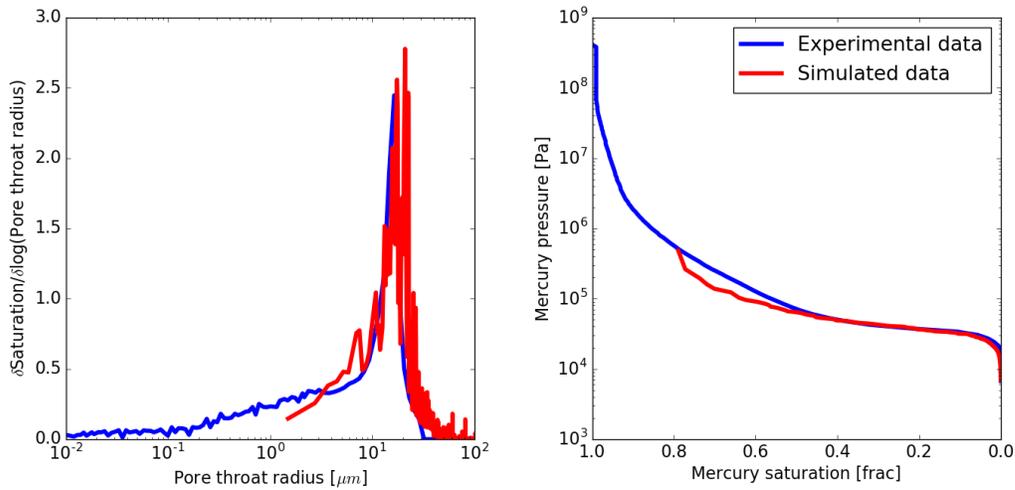

**Figure 3:** Plots of a simulated mercury intrusion for a segmented micro-CT image of a core plug together with experimental mercury intrusion conducted on a sister plug. Note that for the simulated data we do not have pore throats smaller than the image voxel size.

It is evident from Figure 3 that a significant fraction of the pore space is not contained in the resolved porosity in the image. SEM imaging obtains better resolution than micro-CT images. Still pores bellow the image resolution, e.g. pores inside clays, is common when the image capture a REV. As clays are porous, their gray scale values are commonly between the values associated with quartz and porosity, as indicated with a red circle in Figure 2. This enables segmentation of the microporous phase.

### 4.2.1 Effective porosity

The Hagen–Poiseuille equation gives the volumetric flow rate $Q$ through a pipe as

$$Q = \frac{\pi r^4}{8\mu} \nabla P, \qquad (2)$$

where $r$ is the pipe radius, $P$ is the pressure, and $\mu$ is the viscosity. As the flow rate is proportional to the cross-sectional area squared, the smallest pores has little impact on overall flow. If the image resolution is sufficient to properly resolve the pore space containing the backbone of the fluid flow, then simulations on the resolved pore space would capture the important features of the fluid flow. On the other hand, smaller pores give small contribution to the overall flow, and they are therefore of little importance when trying to capture the main impacts on the fluid flow in our simulations. They can however be important for saturations and electrical conductance.

The effective porosity is sometimes defined as the porosity that contributes to fluid flow. As the smallest pores have little effect on the overall fluid flow, they are commonly excluded from the effective porosity. If the image resolution is fine enough for simulations to capture the main portion of the fluid flow, then the resolved porosity corresponds to the effective porosity. Identifying the resolved porosity with effective porosity is thus dependent on sufficient image resolution and quality. The amount of resolved porosity in Figure 3 indicates a sufficient image resolution to capture the main flow paths, as the majority of the pore throats are resolved properly. Tighter rocks would require a higher resolution. In the following we assume that the image resolution is fine enough to resolve the main flow paths, and would therefore equate resolved and effective porosity.

As the contribution to flow from the smallest pores might be ignored, it is expected that the effective porosity has a better correlation with permeability than the total porosity (Nordahl et al., 2014). In Figure 4 we have plotted both helium and effective porosity versus permeability for a set of samples from a field at the Norwegian Continental Shelf (NCS). The thin sections have been produced from end caps of a set of core plugs, and the helium porosity and permeability values are from the corresponding core plug. More information on the imaging procedure can be found in (Lopez et al., 2013).

As expected we observe an improved porosity-permeability correlation for the effective porosity over the helium porosity. On should note that we obtain an improved porosity-permeability correlation despite the fact that the effective porosity is calculated from a thin section representing a cross-section at one end of the core sample. Heterogeneities in the sample not represented at the end-cap cross-section will therefore not be accounted for. Improvements in the porosity-permeability correlation have been observed for a large variety of reservoirs, and the strongest improvements occur for fairly homogeneous reservoirs.

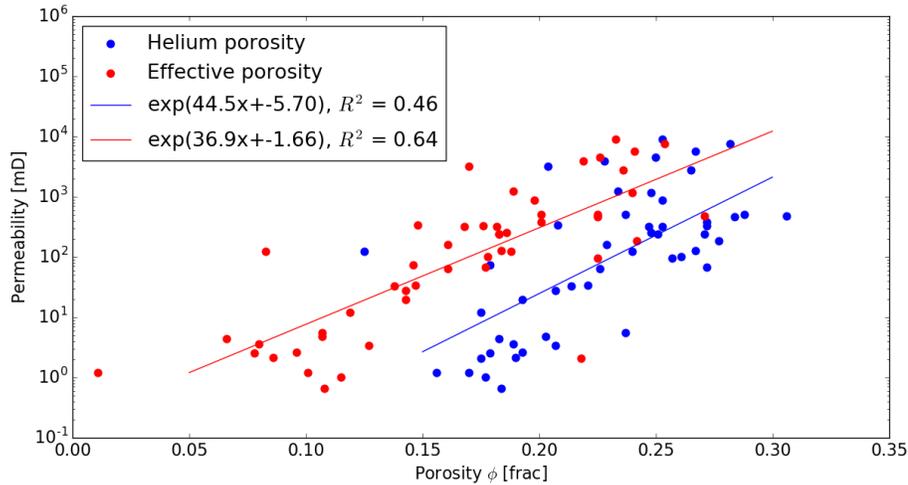

**Figure 4:** Plot comparing the correlation of helium porosity and effective porosity from thin sections versus permeability.

### 4.2.2 Microporosity

While the fluid flow rate in a tube is proportional to the cross-sectional area squared, electrical conductance is proportional to the area. To accurately simulate electrical conductance the unresolved porosity must be accounted for. For lower permeability samples the unresolved porosity is important for simulation of fluid flow too.

For both micro-CT and SEM images unresolved porosity typically gives gray levels between quartz and fully resolved porosity, as observed for the micro-CT image at the left in Figure 6 and for the BSE image in Figure 2. This enables direct segmentation of the unresolved porosity. Let $V_m$ be the fraction of pixels with grey levels between the levels associated with quartz and resolved porosity. Assuming a constant porosity fraction $\mu_m$ for the unresolved porosity, then the amount of porosity associated to the unresolved pores is:

$$\phi_m = \frac{V_m \mu_m}{V_t}, \quad (3)$$

where $V_t$ is the total amount of pixels in the image. If $\phi_e$ is the effective porosity (resolved porosity) in the image, then the total porosity is given by the sum

$$\phi = \phi_e + \phi_m. \quad (4)$$

We can now treat the micro-porosity fraction $\mu_m$ as an adjustable variable to minimize the difference between image porosity and measured porosity. The micro-porosity in samples from the NCS is mainly associated with clays, mostly kaolinite. Depending on clay types, resolution and segmentation, the micro-porosity fraction $\mu_m$ is typically around 0.5 for BSE images of NCS samples with resolution around 1 $\mu m$. For the samples considered in Figure 4 we used a fraction of $\mu_m = 0.4$. Calculated image porosity versus helium porosity for the corresponding plugs is plotted in Figure 5. The clustering around the 1-1 line fraction gives confidence in the microporosity value of $\mu_m = 0.4$.

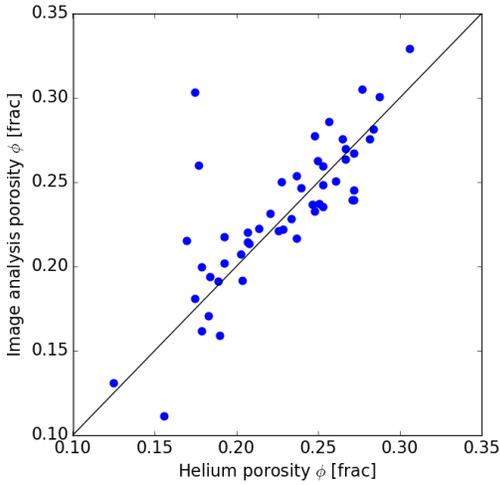

**Figure 5:** Helium porosity versus the calculated image porosity from the corresponding thin sections.

One method to obtain a better porosity image from micro-CT imaging is dry-wet imaging. With this method the same sample is imaged both in dry state and fully saturated with brine (Bhattad et al., 2014; Feali et al., 2012; Long et al., 2013). The brine is added a strongly X-ray attenuating dope to increase the contrast. Example images of dry and wet samples are presented in Figure 6. Such dry-wet imaging has the advantage that it can reveal sub-resolution pores, i.e. microporosity. The red circle in Figure 6 indicates voxels with attenuation between that of pores and quartz. The center image in Figure 6 shows that these voxels still have attenuation values between pore and quartz when pores have a higher X-ray attenuation than quartz as the sample has been filled with doped brine. This indicates that these voxels are influenced by the doped brine, and therefore represents a microporous phase. The difference between the dry and wet image produce an image of the porosity of the sample. The gray-levels of the difference image corresponds to the porosity value for the voxel, and can be utilized to estimate the porosity fraction $\mu_m$ for the unresolved porosity.

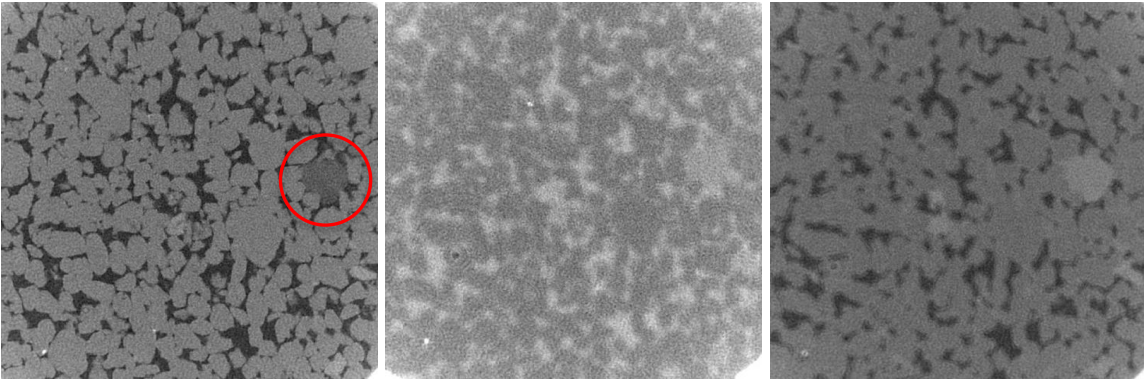

**Figure 6:** Cross-section of micro-CT image of Bentheimer sample. Left image shows dry sample, centre image shows sample saturated with dope brine, while right image show sample when drained to irreducible water saturation. The red circle indicates unresolved porosity.

### 4.2.3 Irreducible water saturation

When oil (equivalently, gas) imbibes a water-filled rock, it first invade the largest pore-throats according to the Young-Laplace equation ( 1 ). Oil will enter smaller and smaller pore-throats as the capillary pressure increase. For a pore-throat size of $0.1 \mu m$, assuming an interfacial tension of $\sigma = 0.03 N/m$ and a contact angle of $\theta = \pi/6$, the Young-Laplace equation gives a capillary pressure of $5.2E5$ Pa, or approximately 5 bar. To reach such a capillary pressure in a reservoir with a water oil system of densities $1000\ kg/m^3$ and $800\ kg/m^3$, respectively, we need an oil column of height $h = 5.2E5 Pa/(200 kg/m^3 * 9.81 m/s^2) = 265 m$, which is considered a high oil column.

Most capillary pressure experiments are stopped at 5 bar, as capillary pressures experienced in the reservoir is assumed to be below this value. When plotting water saturation versus capillary pressure, most sandstones exhibit an asymptote before reaching 5 bar. This asymptote is assumed to distinguish the effective porosity from the micro-porosity, and oil will not enter the micro-porosity. The water saturation at this asymptote is interpreted as the irreducible water saturation $S_{wirr}$. As the capillary pressure curves do not always exhibit an asymptote, the irreducible water saturation is not well defined. It is therefore common to define the irreducible water saturation as *the* water saturation at a capillary pressure of 5 bar.

From the calculations above, an irreducible water saturation defined as the saturation at 5 bar capillary pressure corresponds to a pore-throat size of approximately $0.1\ \mu m$. Given the aspect ratio between pore body and pore throat, this corresponds to porosity not resolved by an image resolution around $1 \mu m$. For such images, we can then estimate the irreducible water saturation with the formula:

$$S_{wirr} = \frac{\phi_m}{\phi}. \tag{5}$$

In Figure 7 we have plotted the irreducible water saturation calculated using Equation ( 5 ) versus permeability. The experiments were conducted using a porous plate method with a maximum end pressure of 5 bar. We observe that the irreducible water saturation from image analysis gave a similar trend as the experiments. For this set of core plugs, most of the unresolved porosity was associated with clays. The negative correlation between irreducible water saturation and permeability is as expected as the clay fill and blocks pores (Coskun and Wardlaw, 1995; Hamon and Pellerin, 1997).

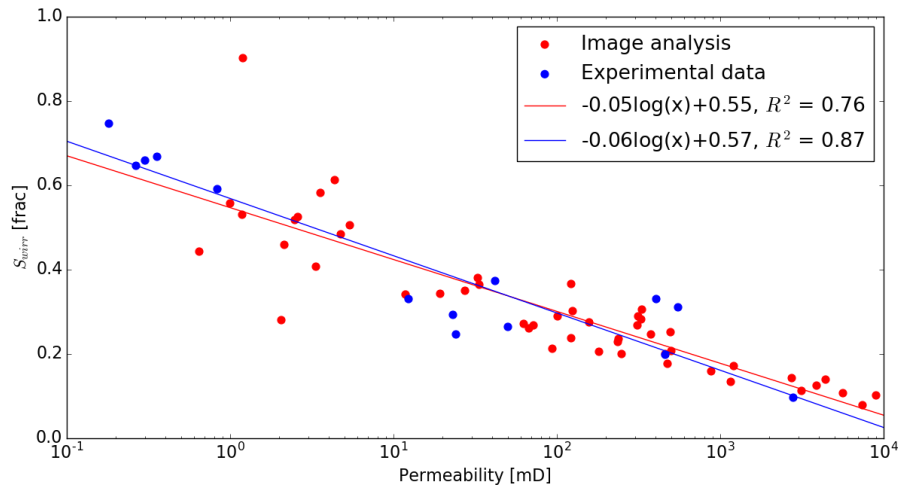

**Figure 7:** Permeability versus irreducible water saturation. The plot includes porous plate experimental data and the calculated values for thin section from the same well.

## 4.3 Grain size and distribution

The grain size distribution is an integral part of the sedimentological description of reservoir rocks. While visual and optical microscopy might give reliable grain size data, one can obtain higher quality data by imaging on the pore scale.

Grain sizes can be extracted directly from a segmented image whenever the image resolves the individual grains. Codes for extracting the grain size are readily available in visualization software such as ImageJ and Avizo. Many grain identification routines are sensitive to noise. Fortunately, the adverse effect from filtering, e.g. a median filter on the segmented binary image, diminishes when the voxel resolution becomes significantly smaller than the grains.

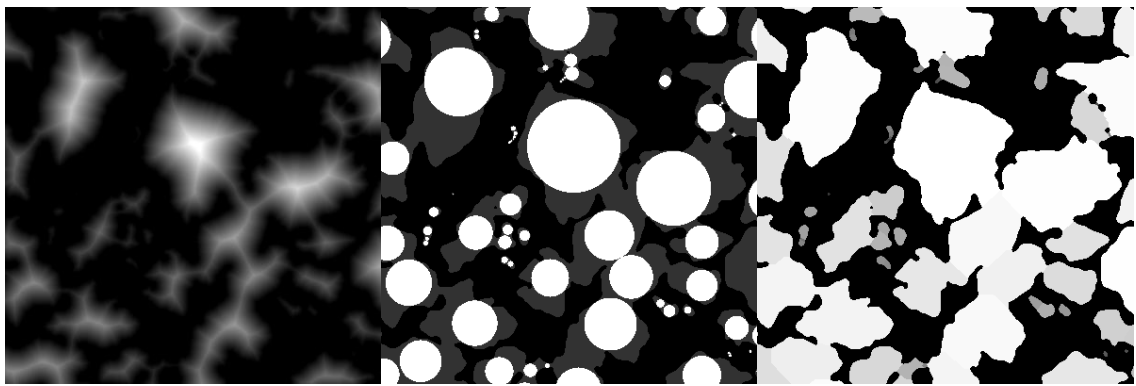

**Figure 8:** Grain size analysis on a binary image, consisting of matrix and porosity. The figure to the left shows the distance to the nearest porosity voxel, the center image shows maximal balls inside the matrix, while the right figure shows the results applying a weighted watershed algorithm starting in the ball-centers.

The procedure for extracting grain sizes from 2D images resembles the procedure for 3D images. An example is shown in Figure 8, where we have applied a maximal ball algorithm on a segmented and filtered BSE image of a thin section of a rock sample. Such statistical measures give an estimate for the grain size distribution, and can be used directly if one only compares 2D distributions. If one need the 3D grain size distribution, e.g. for generating a 3D rock model, the stereological effect needs to be accounted for.

A BSE image of a thin section can be considered as an infinitesimal plane through the rock sample. Given a sphere of radius $r$, a plane cutting through the sphere at a distance $l$ from the center gives a circle of radius $r' = \sqrt{r^2 - l^2}$. This translates into a probability distribution as plotted in Figure 9. The grain size distribution of a monodispersed grain pack is plotted in the same figure. As seen from the plots, there is a fair correspondence between the analytical probability distribution and the disk sizes from the cross-section of the sphere pack.

To back-calculate the original 3D distribution, let us represent the observed size distribution by a vector $\vec{w} = (w_i)$, where $w_0$ corresponds to the smallest grain size bin, while $w_n$ corresponds to the largest. Then let $P = (p_{i,j})$ be the $n \times n$ triangular matrix with

$$p_{i,j} = 1/j \left( \sqrt{j^2 - (i-1)^2} - \sqrt{j^2 - i^2} \right) \forall i \leq j.$$

Then $\sum_{i=0}^{j} p_{i,j} = 1$, and if $\vec{v}$ is the distribution of the 3D sphere radius, then $\vec{w} = P\vec{v}$, thus $\vec{v} = P^{-}\vec{w}$. To ensure $\vec{v}$ only consists of positive elements one can use an iterative method to minimize $\|\vec{w} - P\vec{v}\|$. Note that this estimate gives the size distribution of the spheres represented in the cross-section. To obtain an estimate of the 3D size distribution one need to adjust for the volume the spheres occupy.

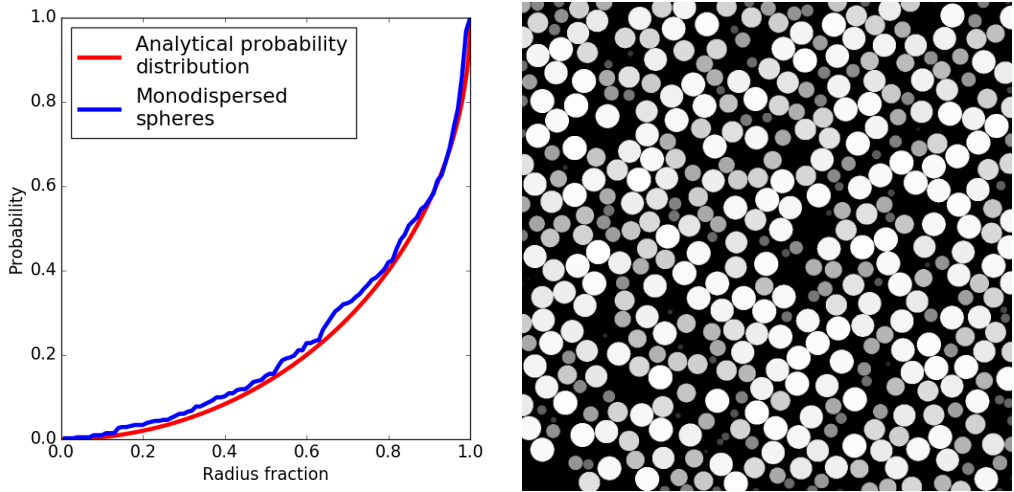

**Figure 9:** Probability distribution for a mono-dispersed sphere pack, together with a cross-section of the sphere pack. The shades of gray in the cross-sectional image correspond to the different disk sizes.

The permeability is described by the pore structure as

$$k = \frac{\phi l_c^2 \tau^2}{C}, \tag{6}$$

where $l_c$ is a characteristic length of the pore space, the tortuosity $\tau<1$ represents the sinuosity of the flow paths, while the constriction factor $C$ represents the aspect ratio between the pore bodies and pore throats (Berg, 2014). The grain size is closely related to the characteristic length $l_c$, and thereby to the permeability.

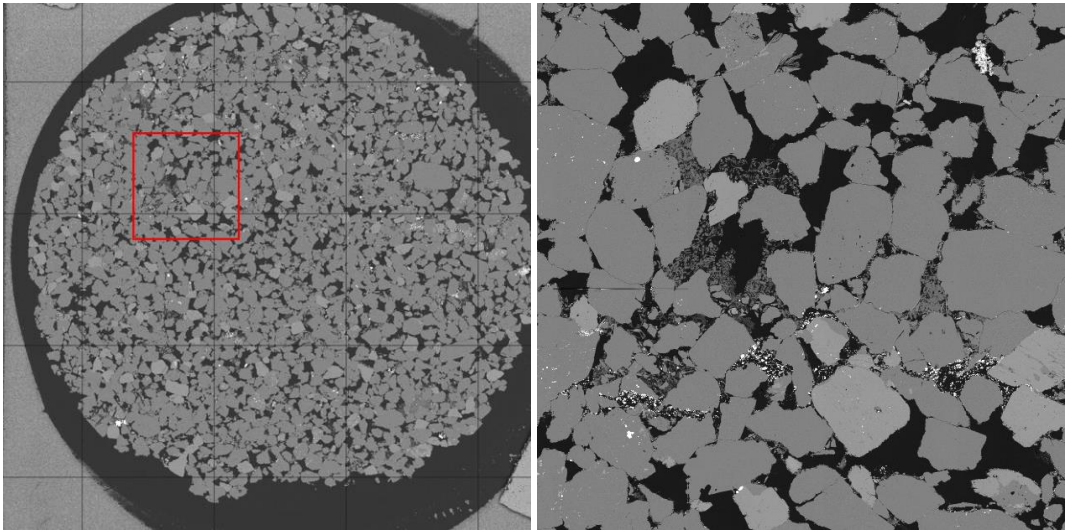

**Figure 10:** Coarse grained sample from Field 1 of 1.5 inch diameter. Image analysis gave an inter-granular porosity of 20.5%, clay fraction of 7.4%, and volume weighted mean grain size of 474$\mu$m. The permeability of the corresponding core plug was 11.8D.

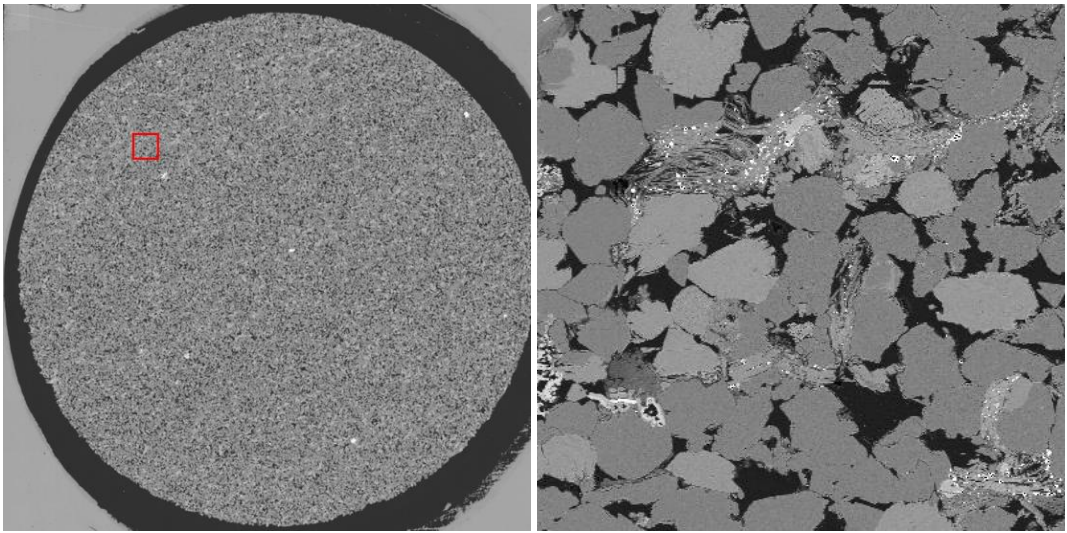

**Figure 11:** Fine grained sample from Field 1 of 1.5 inch diameter. Image analysis gave an inter-granular porosity of 21.4%, clay fraction of 10.6%, and volume weighted mean grain size of 100$\mu$m. The permeability of the corresponding core plug was 477mD.

BSE images of two rock samples from a reservoir at the NCS are shown in Figure 10 and Figure 11. We observe that these samples have clearly different grain sizes. However, when investigating the enlarged section, the grain size distribution and the pore structure has similarities. As the tortuosity $\tau$ and constriction factor $C$ are scale invariant (see (Berg, 2014)), such similarities indicates similar tortuosity and constriction factor for the two samples. Also the porosity for the two samples are similar. Applying Eq. ( 6 ), where subscript 1 is for the coarse sample in Figure 10 while subscript 2 is for the fine grained sample in Figure 11, we obtain:

$$\frac{k_1}{k_2} = \frac{\phi_1 l_1^2 \tau_1^2}{C_1} \frac{C_2}{\phi_2 l_2^2 \tau_2^2} \simeq \frac{l_1^2}{l_2^2} \sim \frac{(474 \mu m)^2}{(100 \mu m)^2} = 22.5$$

This is a fair correspondence with the measured fraction of $k_1/k_2 = 11800/477 = 24.7$, indicating that the grain size is a first order effect for the permeability difference between these two samples.

Grain sorting impacts both the characteristic length, porosity and constriction factor. In general, rocks with a large variance in grain size exhibit a short characteristic length, low porosity and high constriction factor. Large variation in grain size then implies lower permeability from Equation ( 6 ). In Figure 12 we have plotted the mean and standard deviation of the grain size distribution in Krumbein $\phi$-scale. The scatter points are colored by the logarithm of the corresponding permeability divided by porosity. Note that in the Krumbein $\phi$-scale a lower value indicates a larger grain size. Thus a lower mean grain size value in Figure 12 indicates a larger grain size. The larger grain size correlates with a higher value for the permeability divided by porosity, as expected. A lower variance also correlates with higher value for the permeability divided by porosity. We observe that lower values for both mean and variance in Figure 12 correlates with higher permeability divided by porosity.

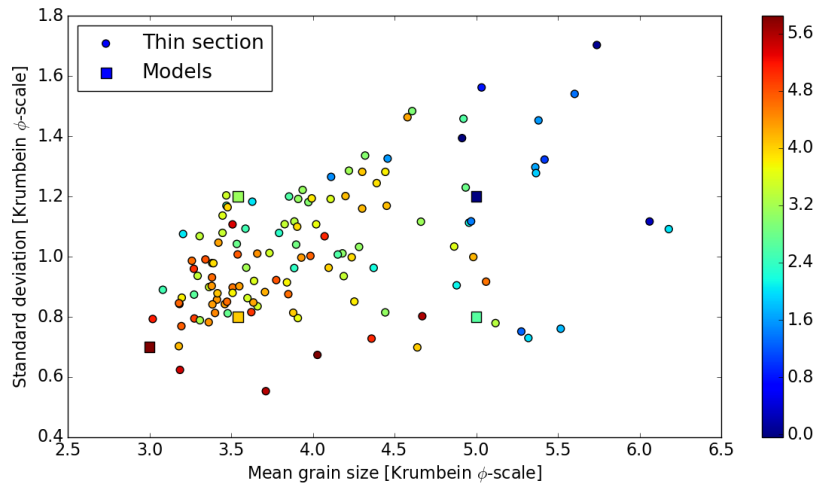

**Figure 12:** Plot of mean and standard deviation of the grain size distribution from a set of thin section data. The plot is colored by the corresponding logarithm of permeability divided by porosity.

In Figure 12 we have also include 5 pore scale models. These models were generated using a grain size distribution given by a normal distribution on the Krumbein $\phi$-scale with the indicated mean and standard deviation. Observe that the permeability divided by porosity for the generated models are in fair correspondence with the experimental data, indicating the importance of the grain size distribution on the permeability.

# 5 Simulation

Digital representation of the pore structure enables the simulation of physical processes to obtain effective porous medium properties. Such effective porous medium properties include absolute and relative permeability, electrical conductance, diffusion and elasticity. In this section we will discuss and compare different methods for calculating these effective properties.

## 5.1 Absolute and relative permeability

Methods to calculate the permeability can be divided into two main groups; modeling directly on grid representation and network modeling (Blunt et al., 2013). Both methods have its strengths and weaknesses.

Direct modeling is usually conducted on a binarized image of the pore structure. Such images can be obtained by the different methods described in Section 3 and 4. Direct modeling thus honors the geometry in the digitized image. On the other hand they are computational demanding and have limited possibilities to include effects happening below the voxel size.

As the name suggest, network modeling extracts a network representation of the pore structure (Fatt, 1956). This honors the underlying topology and overall volumes, however it is a simplification that does not take all the finer details of the pore geometry into account. The main benefit of network modeling is computational efficiency and resolution (Valvatne and Blunt, 2004).

As mentioned in the introduction, capillary dominated pore filling processes during primary drainage has simpler fluid-fluid and fluid-rock interactions. This enables the modeling of these processes solely based on pore structure information (Hilpert and Miller, 2001). Such pore structure based simulations has been shown to closely match experimental pore scale data for primary drainage (Hussain et al., 2014), while imbition processes need more information for accurately modeling (S. Berg et al., 2016). Another option for a capillary dominated displacement is to use the level-set method to track fluid interfaces (Prodanović and Bryant, 2006).

The most popular direct modeling technique is the lattice-Boltzmann method. Historically this method evolve from the lattice-gas automata models, employing a Boltzmann equation to model the single-particle lattice gas (McNamara and Zanetti, 1988). The method relies on a simplified collision and propagation scheme, and can be shown to approximate the Navier-Stokes equation. The popularity of this method partly arises from the fact that it is easy to implementation and well suited for parallelization. In addition it can be extended to multiphase flow as fluid-fluid and fluid-solid interactions are straight forward to include (Ramstad et al., 2012). In particular, phases can easily be tracked as particles represent fluid elements, while there are several different methods for fluid separation (Gunstensen et al., 1991; Mcclure et al., 2016; Shan and Chen, 1993). Fluid separation by both the color gradient and the free energy model has been shown to accurately capture the dynamics of capillary filling events (Ryan T. Armstrong et al., 2016; Zacharoudiou and Boek, 2016).

For slow flow the Navier-Stokes equation can be simplified to the Stoke equation. For single phase flow the Stokes equation has been solved by an artificial compressibility method (Bentz and Martys, 2007; Berg and Held, 2016). Multiphase flow has been treated by solving the Navier-Stokes equation while tracking the fluid interface, e.g. by a volume-of-fluid technique (Hirt and Nichols, 1981). Several groups has used the openFOAM framework for such volume-of-fluid two-phase flow simulations (Ferrari and Lunati, 2013; Raeini et al., 2012).

Another direct modeling technique is the Smoothed Particles Hydrodynamics (SPH), where fluid-fluid surface tension is simulated by interaction forces between different particles (Morris, 2000; Tartakovsky and Meakin, 2006). In contrast to the Lattice-Boltzmann technique, the SPH techniques can handle larger viscosity and density contrasts. The SPH technique has recently been used to model micromodel experiments of multiphase flow (Kunz et al., 2016; Sivanesapillai et al., 2016).

Direct hydrodynamic (DHD) simulation is a fairly new method based on classical continuum hydrodynamics. DHD takes into account thermodynamic energy balance in addition to mass and momentum, while fluid-fluid interaction is modeled with the density functional approach (Demianov et al., 2011). It has been demonstrated that the method can reproduce a wide range of basic two-phase flow mechanisms (R. T. Armstrong et al., 2016). Digital rock modeling applying the DHD method is commercially available (Koroteev et al., 2014).

As mentioned, network modeling involves extracting a network representation of the pore structure, where larger voids defined as pores are connected by smaller voids defined as throats. For complex pore structures there are no unique definition of pores and pore throats, thus different network extraction methods produces different networks (Dong and Blunt, 2009). Reconstructed porous media, as described in Section 3.3, contains information about the borders of the individual grains. Invoking grain information benefits the network extraction in the sense that it reduces the freedom in separating pores and throats (Bakke and Øren, 1997).

The validity of networks could be tested by comparing their transport properties to direct solutions. Even though direct simulations are computational heavier than network simulations, simulation of single phase properties such as permeability and electrical conductance are light enough to be used for quality control of network models. Confidence in the network model is strengthen when calculated values for permeability and electrical conductance are comparable for the direct and network model. Also the individual pore structure descriptors for transport given in Equation ( 6 ) has been applied to validate pore networks (Berg, 2013).

Simulation of multiphase flow in pore networks is typically quasi-static and restricted to a flow regime dominated by capillary forces (Oren et al., 1998). There exists dynamic simulators that can account for viscous forces, however these are still at a research state (Nguyen et al., 2006). The picture is quite opposite for direct modeling; direct modeling takes viscous forces into account, and simulations get more computational demanding when capillary forces starts to dominate. With direct simulations one circumvents the inherent issue of creating a representative network. Unfortunately they are so computational demanding that it is hard to capture a REV and at the same time resolving film flow. Networks have in practice infinite resolution, and film flow is incorporated in commercially available simulators.

The pros and cons for direct modeling and network modeling indicate that the method of choice is highly dependent the objectives for the investigation. Exemplarily, when modeling high flow rates or low surface tension the viscous forces become dominant, and direct modeling is typically the method of choice. Conversely, when modeling end-point saturations in a capillary dominated regime, then film flow is a first order effect and network modeling is typically the method of choice. The service industry delivering digital rock modeling has been divided between focusing on direct modeling (Koroteev et al., 2014) and network modeling (Bakke and Øren, 1997).

## 5.2 Electrical conductance and steady-state diffusion

Both electrical conductance and steady-state diffusion is governed by the Laplace equation, thus they have equivalent solutions. We will only consider electrical conductance in this paper. From the Hagen–Poiseuille equation, Equation ( 2 ), we saw that for a tube the fluid flow rate is proportional to the cross-sectional area squared. However, the electrical conductance of an electrolyte in a tube is correlated with the area. The effective electrical conductance $\sigma_c$ for a non-conductive porous medium filled with an electrolyte of conductance $\sigma$ is described by the pore structure as (Berg, 2012):

$$\frac{\sigma_c}{\sigma} = \frac{\phi \tau_c^2}{C_c}. \quad (7)$$

The equation is analogue to Equation ( 6 ) for permeability, however it does not contain a characteristic length as the electrical conductance is correlated to the area instead of the area square. Note that the tortuosity $\tau_c$ is an explicit integral of electric field line lengths. As the electric field lines are different from the fluid streamlines, the electrical tortuosity $\tau_c$ is different from the fluid flow tortuosity $\tau$. Analogous, the electrical constriction factor $C_c$ differs from the fluid flow constriction factor $C$.

As the electrical conductance correlates with the area, the contribution from the microporosity to overall conductance is significant. Simulation of electrical conductance therefore needs to account for the microporosity. If the microporous phase is assumed homogeneous, then one could calculate its effective electrical conductance on high resolution FIB-SEM images or reconstructed 3D structures from high resolution SEM images of the microporous phase. Such procedures has been conducted for micrite in carbonates and clays in sandstones (Lopez et al., 2012; Ruspini et al., 2016). Without such micro-pore models one could use trend analysis to obtain a relation for the tortuosity and constriction factor versus porosity for the microporous phase (Rezaei-Gomari et al., 2011).

With a 3D model of the pore structure, including effective conductance for the microporous phase, the Laplace equation can be solved using e.g. a conjugate gradient method. The surface conductance can also be included into such calculations (Johnson et al., 1986); however the surface conductance is typically of second order for reservoir salinities. The resulting electric field then gives an effective porous medium conductance $\sigma_c$ for a given electrolyte conductance $\sigma$, which yields the formation factor $F = \sigma/\sigma_c$. For a multiphase flow simulation one might compute the electrical conductance for the water phase for each saturation step, which yields the resistivity index.

## 5.3 Elastic properties

There exists a range of methods for obtaining different elastic properties (Andrä et al., 2013b). Assuming a linear elastic regime, the local equations governing the static elastic behavior can be solved by finite element methods (Garboczi and Day, 1995). Such methods has been applied to grid models of sandstones (Øren et al., 2007). In highly consolidated materials the grain-grain connections are properly resolved. For such consolidated material, e.g. Fountainebleau samples, computed elastic properties is in fair agreement with experimental data (Arns et al., 2002). For less consolidated materials, e.g. sand packs, grain-grain interfaces are not properly resolved. Thus calculation of elastic properties tends to have the same issues with scale and resolution as for fluid flow and electrical properties. Unfortunately, robust workarounds for reservoir sandstones are still not on an industrial standard, thus calculation of elastic properties has not had a wide adoption yet. We will therefore not consider elastic properties in the remainder of this paper.

## 6 Application example

This section describes an example of industrial application of digital rock technology. We will consider a field on the NCS together with a smaller satellite field. The main aim of this digital rock application was to provide flow parameters for the satellite field within a short time frame. Depending on results from the rock characterization, a secondary aim was to apply the generated flow parameters for the main field.

Based on results in (Lopez et al., 2010), we consider the digital rock modeling presented in this section reliable for samples with resolved porosity larger than 10%, permeability larger than 10mD and clay content below 15%. While no objective quantification of grain size distribution could cover all eventualities, in our experience the radius ratio between the smallest and largest grains should not be much larger than 1:10 for models to properly resolve both the smallest and largest grains. This operational range is for model sizes around 1000^3 voxels. Larger models or other modeling techniques can extend the operational range.

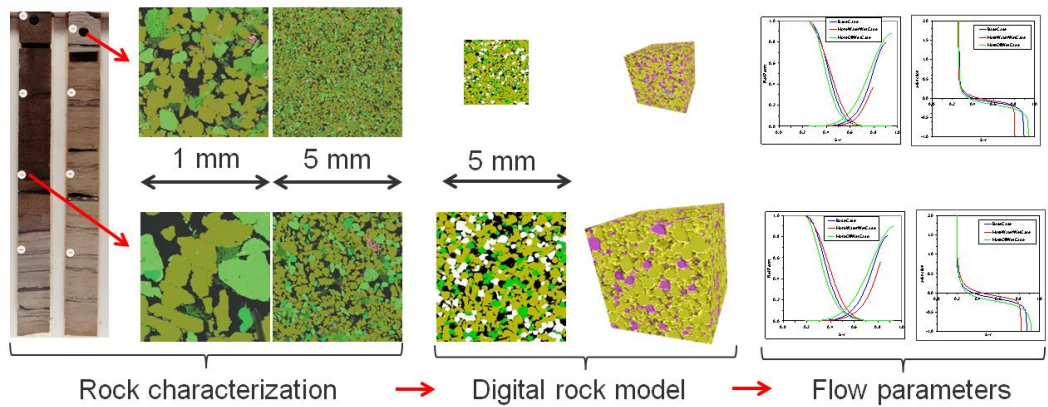

**Figure 13:** A flow chart of the workflow for generating flow parameters by digital rock modeling. The colors for the digital rock models are not directly related to each other, nor to the thin section images.

A flow chart of the applied workflow is illustrated in Figure 13. This workflow builds on an assumption of homogeneity; we assume the existence of rock types with similar cross-property trends. These cross-property trends are further based on the concept of a REV as introduced by Bear (Bear, 1988): We assume the existence of a volume that is large enough to capture a representative amount of the heterogeneity, such that the investigate properties and thereby the cross-property trend becomes independent of the averaging volume. Earlier studies indicate that such a REV concept seems appropriate for porosity and single phase transport properties (Arns et al., 2005; Øren et al., 2007). Higher flow rates (higher capillary number) see the development of more ganglia flow, and the size of a moving ganglia could span many length scales (Armstrong et al., 2014; Georgiadis et al., 2013). For typical reservoir flow rates we assume less ganglia dynamics, thus the two-phase extension of Darcy's law and the REV concept might be valid for most of the saturation range. It should be noted that these issues are shared with conventional SCAL analysis.

The workflow, as illustrated in Figure 13, started with core viewing of samples from both the satellite and the main field. Visual inspection indicated similar geological features in wells from the satellite and the main field. Based on the core viewing and log interpretation the geologist concluded that the wells contained roughly three different rock types, visually classified as fine, medium and coarse sand. The upper part of the different wells started with medium sand interlayered with one or two sections of coarse sand. Below these sand packs was a section of fine sand.

All three rock types consisted of homogeneous sand. As these sands were resolved in the corresponding reservoir models, it was concluded that there was no need for upscaling of flow parameters. Less homogeneous sands would have implied one or several upscaling steps to obtain flow parameters for the appropriate scale (Bhattad et al., 2014; Odsæter et al., 2015; Rustad et al., 2008).

A set of thin sections were chosen to cover the internal spread in properties for the three rock types. Both BSE and EDS images of the thin sections were obtained, with examples shown in Figure 10 and Figure 11. The grain size distributions were extracted as described in subsection 4.3. Resulting grain size distributions are plotted in Figure 14. The naming for the three sand classes are not necessarily in accordance with the Wentworth size terms, the naming is still kept for simplicity.

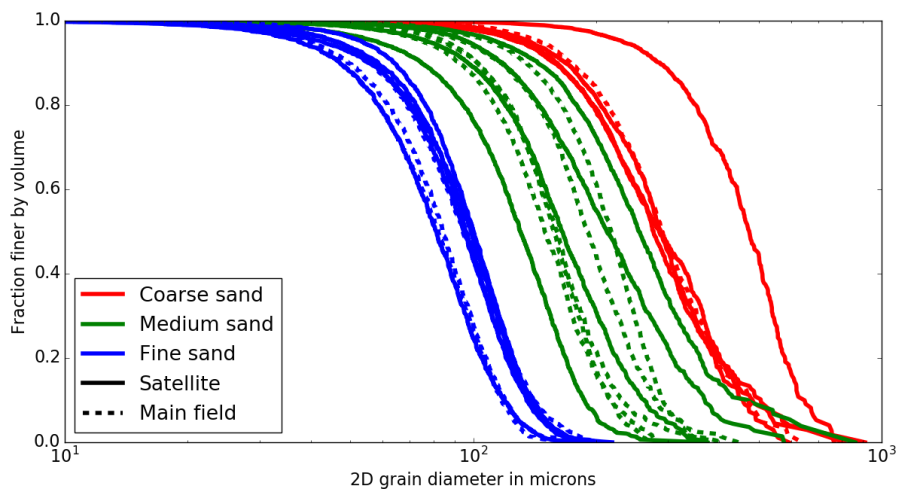

**Figure 14:** Grain size distribution for samples from the satellite and main field. The colors correspond to the associated rock type determined during the core viewing.

The SEM images indicated that most microporosity was associated with clay. Image analysis gave slightly lower amounts of clays for the coarse sand compared to the medium sand. The fine sand had a large variation in clay amount which correlated with depth. The upper samples, close to the coarse and medium sand, had clay levels comparable to the coarse and medium samples, while deeper samples contained significantly more clay.

The differences in grain size and clay amount translate into different trends for permeability versus porosity, as seen in Figure 15. The coarse and medium samples has higher permeability values for comparable porosity values, and the larger spread in clay amount for the fine scale samples translates into a large spread in both permeability and porosity.

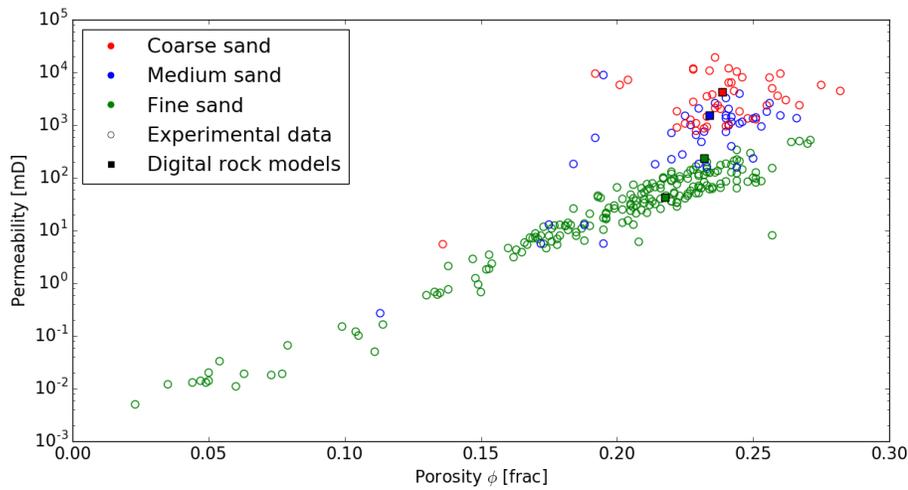

**Figure 15:** Porosity versus permeability for different sands, comparing models to experimental data.

Four digital rock models were created to represent the three different sand types, with an extra model to capture the spread in clay values for the fine grained rock type. Each of these rock models were generated based on statistics from a single thin section. The porosity, grain size distribution and mineralogy values for the four corresponding thin sections were all close to the average for the rock type. Thus the generated digital rock model was assumed to represent the rock type in general, and not just the single thin section from which it was generated. The digital rock models are also included in Figure 15, and we observe that they are all aligned with the permeability trend for the corresponding experimental values for the associated rock type. This is a further confirmation that the digital rock models are representative for their corresponding rock types.

Network representations were extracted from the digital rock models. Fluid-fluid and fluid-rock information for two-phase oil-water simulations was estimated based on experimental data from the main field. This included an estimated value for interfacial tension value and an estimate of the wettability of the samples.

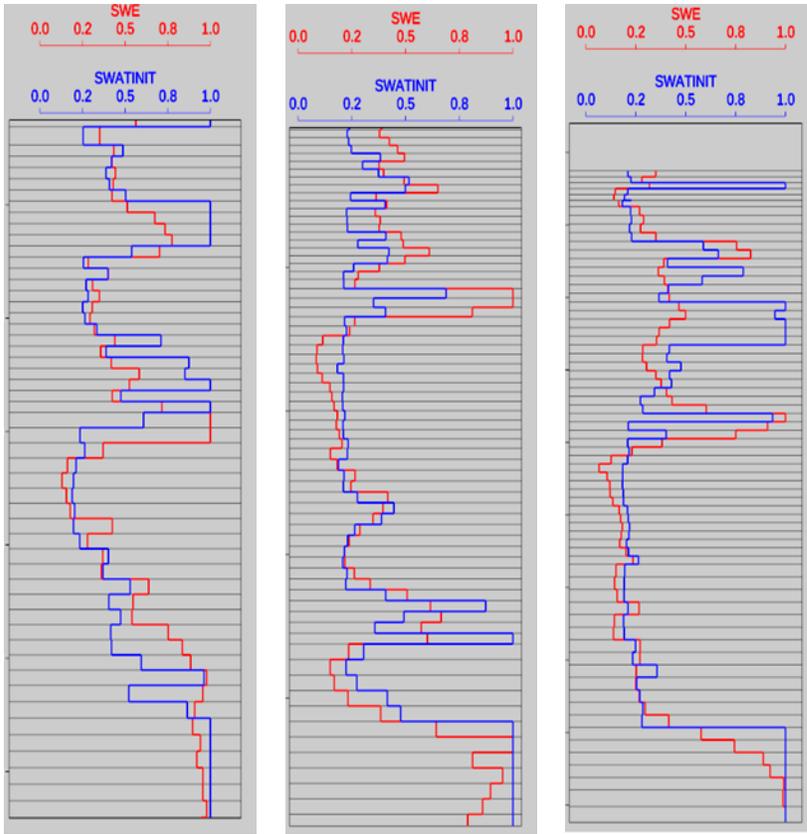

**Figure 16:** In blue are saturations calculated using a single set of Archie parameters and irreducible water saturation value. Computer-processed interpretation (CPI) from well log data is plotted in red. The horizontal gray lines indicate cell boundaries in the corresponding reservoir model, with height in the order of 1.5 meters.

Capillary pressure curves from the digital rock modeling was assigned to facies in a geo-model of the main field. Together with depths for fluid contacts, these capillary pressure curves was used to estimate water saturation in the model. The obtained water-saturation logs are plotted with blue log curves in Figure 16. The same plot includes computer-processed interpretation (CPI) from well log data, displayed with red log curves. There is a clear correlation between the two independent methods, which strengthens the representativeness of the digital rock models even further.

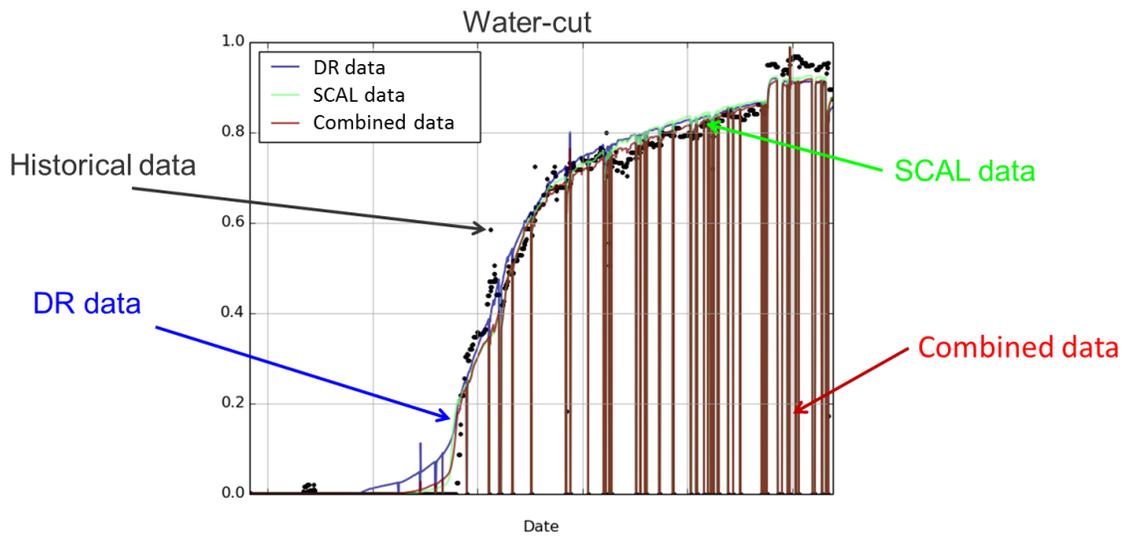

**Figure 17:** Water-cut for a single producer at the main field. This plot contains both historical data and simulation results. The historical data is indicated with black dots, while results from flow simulations are indicated using lines. The blue line is simulations using multiphase flow parameters from digital rock modeling, the green line is simulations using SCAL data, while the red curve shows a simulation using both sets of data.

The resulting two-phase flow parameters were implemented into the dynamic reservoir model for the main field. Given the good results for the water saturation model described above, the digital rock models were assigned to facies in the same way as for the water saturation modeling. The reservoir model with multiphase data from digital rock modeling produced a fair match with historical data, as seen for an exemplary production well in Figure 17. We observe that the simulations based on digital rock modeling results in an earlier water break-through than the historical data. Except for the water-breakthrough, and given that the simulation model has not been history matched, the simulation results are in fair agreement with the historical data.

It took less than two months from the first core viewing to delivery of digital rock modeling data. These data was then obtained significantly faster than conventional experimental data. They were indeed delivered fast enough to be included in simulations for decisions on field development for the satellite field.

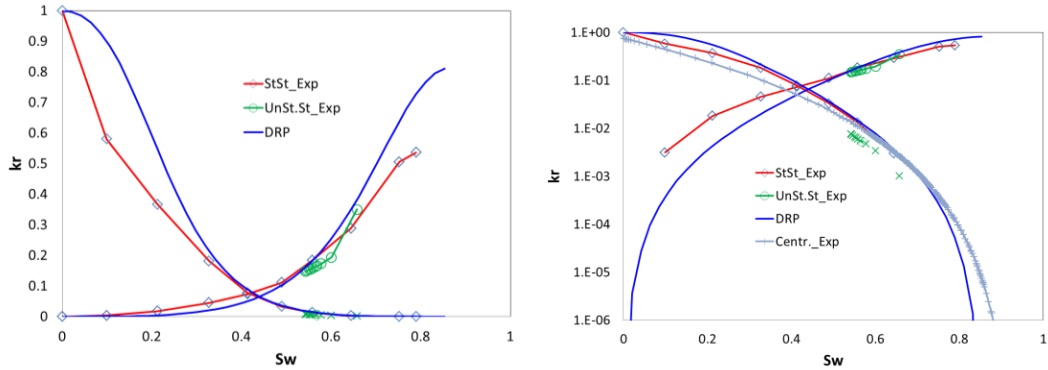

**Figure 18:** Relative permeability curves from digital rock modeling (blue lines) compared to a set of experimental data. The experiments were conducted on sister plugs. For all curves the irreducible water saturation is scaled out for comparison reasons.

When traditional experimental data was obtained at a later stage, the data was compared to the results from digital rock modeling. Comparisons for one sample is included in Figure 18. This comparison shows clear differences for the oil relative permeability curve at low water saturations. There are also significant differences for the endpoints, with the steady-state and unsteady state flooding experiments indicating higher residual oil values than the digital rock modeling, while the centrifuge indicates a slightly lower value. The water curves seems to follow similar trends toward the residual oil endpoint. Given the spread between different experimental methods, there is a fair agreement between digital rock models and experimental data overall.

The SCAL data was implemented into the reservoir model considered above. This produced a significantly better result for the water break-through, and arguably better results for the remainder of the production curve, as shown in Figure 17. It was observed that the main difference between the digital rock data and the SCAL data was that the digital rock data lacked a representative curve for the lowest permeability saturation region. We therefore tried to augment the digital rock data with a single SCAL relative permeability curve for the low permeability saturation region. The simulation result is denoted combined data in Figure 17. We observe that the combined data gave an improved match over the SCAL data.

## 7 Summary

In this paper we have presented an overview of digital rock technology from an industrial perspective. We have provided examples of industrial employment of digital rock technology, and also pointed to the adoption of such technology.

While imaging is a starting point for digital rock simulations, it is also a means in itself. Images of rock samples have direct application in formation evaluation, rock typing and estimation of irreducible water saturation, as demonstrated in this paper.

As with imaging, also rock modeling is a means in itself. In contrast with both 3D imaging and traditional experiments, rock modeling gives the ability to do sensitivity studies by constructing 3D models where only one property is changed. This was exemplified by creating rock models with different grain size distribution, which revealed correspondence between grain size distribution and permeability. Investigations of the individual effects of changing porosity, clay amount, degree of cementation, etc. could be performed similarly.

It is fairly straightforward to simulate the fluid permeability given a 3D image with sufficient resolution to resolve the backbone of the fluid flow and at the same time large enough to capture a REV. For two-phase flow properties the situation is more complex. Network modeling has the advantage of practically infinite resolution and computational efficiency. As films are resolved in network models, they might be more reliable for obtaining the residual oil saturation. On the other hand, simulations directly on the grid remove the inherent uncertainty in network extraction. Including viscous forces is more straightforward for direct gird simulations than in network simulations, therefore direct grid simulations are typically the preferred tool for simulation of enhanced recovery processes.

Industrial applications were exemplified by an application to a field at the NCS. The direct relation between simulation and imaging offers a robust link between rock types and flow parameters. The imaging part of digital rock technology contributes to the rock typing, at the same time simulation provides petrophysical and flow parameters directly related to the images and thereby the rock types. This helps bridge the gap between geological interpretation, petrophysics and reservoir simulations.

In the digital rock study described in this paper, flow parameters from digital rock technology was obtained in a timeframe of 2 month, which was a fraction of the time used for traditional SCAL. This exemplifies how digital rock technology can be a tool for fast track decisions. The study also shows how digital rock technology can extend traditional experiments, as traditional experiments and digital rock technology provides mutual benefits. Imaging and modeling links geological interpretation and transport properties, and demonstrates how traditional experiments benefits from digital rock technology. Another example is the possibility to do sensitivity studies. On the other hand, fast experiments such as mercury injection are important for calibration of digital rock models.

Digital rock imaging, modeling and simulation have had a large present in academia for decades. In the last decade it has moved from a fringe activity to a mature technology for the oil industry. Most large oil companies now have in-house tools and development efforts, while smaller companies are more dependent on a number of service providers who deliver digital rock technology.

# 8  Acknowledgment

We thank Statoil for permission to publish this work.

# 9  Bibliography

Aarnes, J.E., Kippe, V., Lie, K.-A., Rustad, A.B., 2007. Modelling of multiscale structures in flow simulations for petroleum reservoirs, in: Geometric Modelling, Numerical Simulation, and Optimization. Springer, pp. 307–360.
Adler, P.M., Jacquin, C.G., Thovert, J.-F., 1992. The formation factor of reconstructed porous media. Water Resour. Res. 28, 1571–1576.

Al-Raoush, R., Papadopoulos, A., 2010. Representative elementary volume analysis of porous media using X-ray computed tomography. Powder Technol. 200, 69–77. https://doi.org/10.1016/j.powtec.2010.02.011

Andersen, M., Duncan, B., McLin, R., 2013. Core truth in formation evaluation. Oilfield Rev. 25, 16–25.

Andrä, H., Combaret, N., Dvorkin, J., Glatt, E., Han, J., Kabel, M., Keehm, Y., Krzikalla, F., Lee, M., Madonna, C., 2013a. Digital rock physics benchmarks—Part I: Imaging and segmentation. Comput. Geosci. 50, 25–32.

Andrä, H., Combaret, N., Dvorkin, J., Glatt, E., Han, J., Kabel, M., Keehm, Y., Krzikalla, F., Lee, M., Madonna, C., 2013b. Digital rock physics benchmarks—Part II: Computing effective properties. Comput. Geosci. 50, 33–43.

Andrew, M., Bijeljic, B., Blunt, M.J., 2014. Pore-by-pore capillary pressure measurements using X-ray microtomography at reservoir conditions: Curvature, snap-off, and remobilization of residual CO2. Water Resour. Res. 50, 8760–8774.

Archie, G.E., 1942. The electrical resistivity log as an aid in determining some reservoir characteristics. Trans. AIME 146, 54–62.

Armstrong, R. T., Berg, S., Dinariev, O., Evseev, N., Klemin, D., Koroteev, D., Safonov, S., 2016. Modeling of Pore-Scale Two-Phase Phenomena Using Density Functional Hydrodynamics. Transp. Porous Media 112, 577–607.

Armstrong, R.T., Georgiadis, A., Ott, H., Klemin, D., Berg, S., 2014. Critical capillary number: Desaturation studied with fast X-ray computed microtomography. Geophys. Res. Lett. 41, 55–60.

Armstrong, Ryan T., McClure, J.E., Berrill, M.A., Rücker, M., Schlüter, S., Berg, S., 2016. Beyond Darcy's law: The role of phase topology and ganglion dynamics for two-fluid flow. Phys. Rev. E 94, 043113.

Arns, C.H., Knackstedt, M.A., Martys, N.S., 2005. Cross-property correlations and permeability estimation in sandstone. Phys. Rev. E 72, 046304. https://doi.org/10.1103/PhysRevE.72.046304

Arns, C.H., Knackstedt, M.A., Pinczewski, W.V., 2001. Accurate estimation of transport properties from microtomographic images. Geophys. Res. Lett. 28, 3361–3364.

Arns, C.H., Knackstedt, M.A., Pinczewski, W.V., Garboczi, E.J., 2002. Computation of linear elastic properties from microtomographic images: Methodology and agreement between theory and experiment. Geophysics 67, 1396–1405.

Bakke, S., Øren, P.-E., 1997. 3-D pore-scale modelling of sandstones and flow simulations in the pore networks. Spe J. 2, 136–149.

Bear, J., 1988. Dynamics of fluids in porous media. Dover.

Bekri, S., Howard, J., Muller, J., Adler, P.M., 2003. Electrical Resistivity Index in Multiphase Flow through Porous Media. Transp. Porous Media 51, 41–65. https://doi.org/10.1023/A:1021229106005

Bentz, D.P., Martys, N.S., 2007. A stokes permeability solver for three-dimensional porous media. US Department of Commerce, Technology Administration, National Institute of Standards and Technology.

Berg, C.F., 2014. Permeability description by characteristic length, tortuosity, constriction and porosity. Transp. Porous Media 103, 381–400.

Berg, C.F., 2013. A Microstructure Description of the Relationship between Formation Resistivity Factor and Porosity.

Berg, C.F., 2012. Re-examining Archie's law: Conductance description by tortuosity and constriction. Phys. Rev. E 86, 046314.
Berg, C.F., Held, R., 2016. Fundamental Transport Property Relations in Porous Media Incorporating Detailed Pore Structure Description. Transp. Porous Media 112, 467–487.
Berg, C.F., Herrick, D., Kennedy, D., 2016. Geometrical Factor of Conductivity in Rocks: Bringing New Rigor to a Mature Model, in: SPWLA 57th Annual Logging Symposium. Society of Petrophysicists and Well-Log Analysts.
Berg, S., Ott, H., Klapp, S.A., Schwing, A., Neiteler, R., Brussee, N., Makurat, A., Leu, L., Enzmann, F., Schwarz, J.-O., 2013. Real-time 3D imaging of Haines jumps in porous media flow. Proc. Natl. Acad. Sci. 110, 3755–3759.
Berg, S., Rücker, M., Ott, H., Georgiadis, A., van der Linde, H., Enzmann, F., Kersten, M., Armstrong, R.T., de With, S., Becker, J., 2016. Connected pathway relative permeability from pore-scale imaging of imbibition. Adv. Water Resour. 90, 24–35.
Bhattad, P., Young, B., Berg, C.F., Rustad, A.B., Lopez, O., 2014. X-ray Micro-CT Assisted Drainage Rock Typing for Characterization of Flow Behaviour of Laminated Sandstone Reservoirs.
Biswal, B., Øren, P.-E., Held, R.J., Bakke, S., Hilfer, R., 2009. Modeling of multiscale porous media. Image Anal. Stereol. 28, 23–34.
Blunt, M.J., 1998. Physically-based network modeling of multiphase flow in intermediate-wet porous media. J. Pet. Sci. Eng. 20, 117–125.
Blunt, M.J., Bijeljic, B., Dong, H., Gharbi, O., Iglauer, S., Mostaghimi, P., Paluszny, A., Pentland, C., 2013. Pore-scale imaging and modelling. Adv. Water Resour. 51, 197–216.
Bultreys, T., Boone, M.A., Boone, M.N., De Schryver, T., Masschaele, B., Van Hoorebeke, L., Cnudde, V., 2015. Fast laboratory-based micro-computed tomography for pore-scale research: illustrative experiments and perspectives on the future. Adv. Water Resour.
Carman, P.C., 1937. Fluid flow through granular beds. Trans.-Inst. Chem. Eng. 15, 150–166.
Coskun, S.B., Wardlaw, N.C., 1995. Influences of pore geometry, porosity and permeability on initial water saturation—An empirical method for estimating initial water saturation by image analysis. J. Pet. Sci. Eng. 12, 295–308.
Cromwell, V., Kortum, D.J., Bradley, D.J., 1984. The use of a medical computer tomography (CT) system to observe multiphase flow in porous media, in: SPE Annual Technical Conference and Exhibition. Society of Petroleum Engineers.
Curtis, M.E., Ambrose, R.J., Sondergeld, C.H., 2010. Structural Characterization of Gas Shales on the Micro- and Nano-Scales. Society of Petroleum Engineers. https://doi.org/10.2118/137693-MS
Demianov, A., Dinariev, O., Evseev, N., 2011. Density functional modelling in multiphase compositional hydrodynamics. Can. J. Chem. Eng. 89, 206–226.
Dilks, A., Graham, S.C., 1985. Quantitative mineralogical characterization of sandstones by back-scattered electron image analysis. J. Sediment. Res. 55.
Dong, H., Blunt, M.J., 2009. Pore-network extraction from micro-computerized-tomography images. Phys. Rev. E 80, 036307.

Fatt, I., 1956. The network model of porous media.

Feali, M., Pinczewski, W., Cinar, Y., Arns, C.H., Arns, J.-Y., Francois, N., Turner, M.L., Senden, T., Knackstedt, M.A., 2012. Qualitative and quantitative analyses of the three-phase distribution of oil, water, and gas in bentheimer sandstone by use of micro-CT imaging. SPE Reserv. Eval. Eng. 15, 706–711.

Ferrari, A., Lunati, I., 2013. Direct numerical simulations of interface dynamics to link capillary pressure and total surface energy. Adv. Water Resour. 57, 19–31.

Flannery, B.P., Deckman, H.W., Roberge, W.G., D'AMICO, K.L., 1987. Three-dimensional X-ray microtomography. Science 237, 1439–1444.

Fredrich, J.T., Lakshtanov, D.L., Lane, N.M., Liu, E.B., Natarajan, C.S., Ni, D.M., Toms, J.J., 2014. Digital Rocks: Developing An Emerging Technology Through To A Proven Capability Deployed In The Business. Society of Petroleum Engineers. https://doi.org/10.2118/170752-MS

Garboczi, E.J., Day, A.R., 1995. An algorithm for computing the effective linear elastic properties of heterogeneous materials: three-dimensional results for composites with equal phase Poisson ratios. J. Mech. Phys. Solids 43, 1349–1362.

Georgiadis, A., Berg, S., Makurat, A., Maitland, G., Ott, H., 2013. Pore-scale micro-computed-tomography imaging: Nonwetting-phase cluster-size distribution during drainage and imbibition. Phys. Rev. E 88, 033002.

Gilliland, R.E., Coles, M.E., 1990. Use of CT scanning in the investigation of damage to unconsolidated cores, in: SPE Formation Damage Control Symposium. Society of Petroleum Engineers.

Gomari, K.A.R., Berg, C.F., Mock, A., Øren, P. al-Eric, Petersen Jr, E., Rustad, A., Lopez, O., 2011. Electrical and petrophysical properties of siliciclastic reservoir rocks from pore-scale modeling, in: SCA2011-20 Presented at the 2011 SCA International Symposium, Austin, Texas.

Gottlieb, P., Wilkie, G., Sutherland, D., Ho-Tun, E., Suthers, S., Perera, K., Jenkins, B., Spencer, S., Butcher, A., Rayner, J., 2000. Using quantitative electron microscopy for process mineralogy applications. JoM 52, 24–25.

Grader, A., Kalam, M., Toelke, J., Mu, Y., Derzhi, N., Baldwin, C., Stenger, B., 2010. A comparative study of digital rock physics and laboratory SCAL evaluations of carbonate cores. SCA2010-24 Novia Scotia.

Grove, C., Jerram, D.A., 2011. jPOR: An ImageJ macro to quantify total optical porosity from blue-stained thin sections. Comput. Geosci. 37, 1850–1859.

Gunstensen, A.K., Rothman, D.H., Zaleski, S., Zanetti, G., 1991. Lattice Boltzmann model of immiscible fluids. Phys. Rev. A 43, 4320.

Hamon, G., Pellerin, F.M., 1997. Evidencing capillary pressure and relative permeability trends for reservoir simulation, in: SPE Annual Technical Conference and Exhibition. Society of Petroleum Engineers.

Hilfer, R., Manwart, C., 2001. Permeability and conductivity for reconstruction models of porous media. Phys. Rev. E 64, 021304.

Hilfer, R., Zauner, T., Lemmer, A., Biswal, B., 2015. Threedimensional microstructures at submicron resolution.

Hilpert, M., Miller, C.T., 2001. Pore-morphology-based simulation of drainage in totally wetting porous media. Adv. Water Resour. 24, 243–255.

Hirt, C.W., Nichols, B.D., 1981. Volume of fluid (VOF) method for the dynamics of free boundaries. J. Comput. Phys. 39, 201–225.

Honarpour, M.M., Cromwell, V., Hatton, D., Satchwell, R., 1985. Reservoir rock descriptions using computed tomography (CT), in: SPE Annual Technical Conference and Exhibition. Society of Petroleum Engineers.

Hove, A.O., Ringen, J.K., Read, P.A., 1987. Visualization of laboratory corefloods with the aid of computerized tomography of X-rays. SPE Reserv. Eng. 2, 148–154.

Hurst, A., Nadeau, P.H., 1995. Clay microporosity in reservoir sandstones: an application of quantitative electron microscopy in petrophysical evaluation. AAPG Bull. 79, 563–573.

Hussain, F., Pinczewski, W.V., Cinar, Y., Arns, J.-Y., Arns, C.H., Turner, M.L., 2014. Computation of relative permeability from imaged fluid distributions at the pore scale. Transp. Porous Media 104, 91–107.

Iassonov, P., Gebrenegus, T., Tuller, M., 2009. Segmentation of X-ray computed tomography images of porous materials: A crucial step for characterization and quantitative analysis of pore structures. Water Resour. Res. 45.

Johnson, D.L., Koplik, J., Schwartz, L.M., 1986. New pore-size parameter characterizing transport in porous media. Phys. Rev. Lett. 57, 2564.

Kalam, Z., Seraj, S., Bhatti, Z., Mock, A., Oren, P.E., Ravlo, V., Lopez, O., 2012. Relative permeability assessment in a giant carbonate reservoir using Digital Rock Physics, in: SCA2012-03, International Symposium, Aberdeen, United Kingdom.

Kelly, S., El-Sobky, H., Torres-Verdín, C., Balhoff, M.T., 2016. Assessing the utility of FIB-SEM images for shale digital rock physics. Adv. Water Resour. 95, 302–316.

Knackstedt, M.A., Jaime, P., Butcher, A., Botha, P., Middleton, J., Sok, R., 2010. Integrating Reservoir Characterization: 3D Dynamic, Petrophysical and Geological Description of Reservoir Facies. Society of Petroleum Engineers. https://doi.org/10.2118/133981-MS

Koroteev, D., Dinariev, O., Evseev, N., Klemin, D., Nadeev, A., Safonov, S., Gurpinar, O., Berg, S., van Kruijsdijk, C., Armstrong, R., Myers, M.T., Hathon, L., de Jong, H., 2014. Direct Hydrodynamic Simulation of Multiphase Flow in Porous Rock. Petrophysics 55, 294–303.

Kozeny, J., 1927. Über kapillare Leitung des Wassers im Boden:(Aufstieg, Versickerung und Anwendung auf die Bewässerung). Hölder-Pichler-Tempsky.

Kunz, P., Zarikos, I.M., Karadimitriou, N.K., Huber, M., Nieken, U., Hassanizadeh, S.M., 2016. Study of multi-phase flow in porous media: comparison of sph simulations with micro-model experiments. Transp. Porous Media 114, 581–600.

Latief, F.D.E., Biswal, B., Fauzi, U., Hilfer, R., 2010. Continuum reconstruction of the pore scale microstructure for Fontainebleau sandstone. Phys. Stat. Mech. Its Appl. 389, 1607–1618.

Lemmens, H., Butcher, A., Botha, P.W., 2010. FIB/SEM and Automated Mineralogy for Core and Cuttings Analysis. Society of Petroleum Engineers. https://doi.org/10.2118/136327-MS

Leu, L., Berg, S., Enzmann, F., Armstrong, R.T., Kersten, M., 2014. Fast X-ray micro-tomography of multiphase flow in berea sandstone: A sensitivity study on image processing. Transp. Porous Media 105, 451–469.

Liu, Z., Herring, A., Arns, C., Berg, S., Armstrong, R.T., 2017. Pore-Scale Characterization of Two-Phase Flow Using Integral Geometry. Transp. Porous Media 1–19.

Lohne, A., Virnovsky, G.A., Durlofsky, L.J., 2006. Two-stage upscaling of two-phase flow: from core to simulation scale. SPE J. 11, 304–316.

Long, H., Nardi, C., Idowu, N., Carnerup, A., Øren, P.E., Knackstedt, M., Varslot, T., Sok, R.M., Lithicon, A., 2013. Multi-scale imaging and modeling workflow to capture and characterize microporosity in sandstone, in: International Symposium of the Society of Core Analysts, Napa Valley, California, USA.

Lopez, O., Berg, C.F., Bakke, S., Boassen, T., Rustad, A.B., 2013. Thin Section Image Analysis for Estimation of Initial Water Saturation in Siliciclastic Reservoir Rocks.

Lopez, O., Berg, C.F., Rennan, L., Digranes, G., Forest, T., Krisoffersen, A., Bøklepp, B.R., 2016. Quick Core Assessment from CT Imaging: From Petrophysical Properties to Log Evaluation. Int. Symp. Soc. Core Anal. Snowmass Colo. USA.

Lopez, O., Mock, A., Øren, P.E., Long, H., Kalam, Z., Vahrenkamp, V., Gibrata, M., Seraj, S., Chacko, S., Al Hammadi, M., 2012. Validation of fundamental carbonate reservoir core properties using digital rock physics. SCA Aberd. Scotl.

Lopez, O., Mock, A., Skretting, J., Petersen Jr, E.B., Øren, P.-E., Rustad, A.B., 2010. Investigation into the reliability of predictive pore-scale modeling for siliciclastic reservoir rocks, in: SCA2010-41 Presented at the 2010 SCA International Symposium, Halifax, Canada.

Mcclure, J.E., Berrill, M.A., Gray, W.G., Miller, C.T., 2016. Tracking interface and common curve dynamics for two-fluid flow in porous media. J. Fluid Mech. 796, 211–232.

McNamara, G.R., Zanetti, G., 1988. Use of the Boltzmann equation to simulate lattice-gas automata. Phys. Rev. Lett. 61, 2332.

McPhee, C., Reed, J., Zubizarreta, I., 2015. Core Analysis: A Best Practice Guide. Elsevier.

Minnis, M.M., 1984. An automatic point-counting method for mineralogical assessment. AAPG Bull. 68, 744–752.

Morris, J.P., 2000. Simulating surface tension with smoothed particle hydrodynamics. Int. J. Numer. Methods Fluids 33, 333–353.

Mosser, L., Dubrule, O., Blunt, M.J., 2017. Reconstruction of three-dimensional porous media using generative adversarial neural networks. ArXiv Prepr. ArXiv170403225.

Nguyen, V.H., Sheppard, A.P., Knackstedt, M.A., Val Pinczewski, W., 2006. The effect of displacement rate on imbibition relative permeability and residual saturation. J. Pet. Sci. Eng., Reservoir Wettability8th International Symposium on Reservoir Wettability 52, 54–70. https://doi.org/10.1016/j.petrol.2006.03.020

Nordahl, K., Messina, C., Berland, H., Rustad, A.B., Rimstad, E., 2014. Impact of multiscale modelling on predicted porosity and permeability distributions in the fluvial deposits of the Upper Lunde Member (Snorre Field, Norwegian Continental Shelf). Geol. Soc. Lond. Spec. Publ. 387, 85–109.

Nordahl, K., Ringrose, P.S., 2008. Identifying the representative elementary volume for permeability in heterolithic deposits using numerical rock models. Math. Geosci. 40, 753–771.
Odsæter, L.H., Berg, C.F., Rustad, A.B., 2015. Rate Dependency in Steady-State Upscaling. Transp. Porous Media 110, 565–589.
Oh, W., Lindquist, B., 1999. Image thresholding by indicator kriging. IEEE Trans. Pattern Anal. Mach. Intell. 21, 590–602.
Okabe, H., Blunt, M.J., 2004. Prediction of permeability for porous media reconstructed using multiple-point statistics. Phys. Rev. E 70, 066135.
Øren, P.-E., Bakke, S., 2002. Process based reconstruction of sandstones and prediction of transport properties. Transp. Porous Media 46, 311–343.
Øren, P.-E., Bakke, S., Arntzen, O.J., 1998. Extending predictive capabilities to network models. SPE J. 3, 324–336.
Oren, P.-E., Bakke, S., Arntzen, O.J., 1998. Extending predictive capabilities to network models. SPE J. 3, 324–336.
Øren, P.-E., Bakke, S., Held, R., 2007. Direct pore-scale computation of material and transport properties for North Sea reservoir rocks. Water Resour. Res. 43.
Pilotti, M., 2000. Reconstruction of clastic porous media. Transp. Porous Media 41, 359–364.
Pittman, E.D., 1979. Porosity diagenesis and productive capability of sandstone reservoirs.
Prodanović, M., Bryant, S.L., 2006. A level set method for determining critical curvatures for drainage and imbibition. J. Colloid Interface Sci. 304, 442–458. https://doi.org/10.1016/j.jcis.2006.08.048
Purcell, W.R., 1949. Capillary pressures-their measurement using mercury and the calculation of permeability therefrom. J. Pet. Technol. 1, 39–48.
Raeini, A.Q., Blunt, M.J., Bijeljic, B., 2012. Modelling two-phase flow in porous media at the pore scale using the volume-of-fluid method. J. Comput. Phys. 231, 5653–5668.
Ramstad, T., Idowu, N., Nardi, C., Øren, P.-E., 2012. Relative permeability calculations from two-phase flow simulations directly on digital images of porous rocks. Transp. Porous Media 94, 487–504.
Ramstad, T., Øren, P.-E., Bakke, S., 2010. Simulation of two-phase flow in reservoir rocks using a lattice Boltzmann method. SPE J. 15, 917–927.
Rezaei-Gomari, S., Berg, F., Mock, A., Øren, P.E., Petersen Jr, E.B., Rustad, A.B., Lopez, O., 2011. Electrical and petrophysical properties of siliciclastic reservoir rocks from pore-scale modeling.
Ruspini, L.C., Lindkvist, G., Bakke, S., Alberts, L., Carnerup, A.M., Øren, P.E., 2016. A Multi-Scale Imaging and Modeling Workflow for Tight Rocks. Society of Petroleum Engineers. https://doi.org/10.2118/180268-MS
Rustad, A.B., Theting, T.G., Held, R.J., 2008. Pore space estimation, upscaling and uncertainty modelling for multiphase properties, in: SPE Symposium on Improved Oil Recovery. Society of Petroleum Engineers.
Schembre-McCabe, J., Salazar-Tio, R., Ball, G., Kamath, J., n.d. A Framework to Validate Digital Rock Technology. SCA2011-28.

Schwartz, L.M., Auzerais, F., Dunsmuir, J., Martys, N., Bentz, D.P., Torquato, S., 1994. Transport and diffusion in three-dimensional composite media. Phys. Stat. Mech. Its Appl. 207, 28–36.

Schwartz, L.M., Martys, N., Bentz, D.P., Garboczi, E.J., Torquato, S., 1993. Cross-property relations and permeability estimation in model porous media. Phys. Rev. E 48, 4584.

Shah, S.M., Gray, F., Crawshaw, J.P., Boek, E.S., n.d. Micro-computed tomography pore-scale study of flow in porous media: Effect of voxel resolution. Adv. Water Resour. https://doi.org/10.1016/j.advwatres.2015.07.012

Shan, X., Chen, H., 1993. Lattice Boltzmann model for simulating flows with multiple phases and components. Phys. Rev. E 47, 1815.

Sheppard, A.P., Sok, R.M., Averdunk, H., 2004. Techniques for image enhancement and segmentation of tomographic images of porous materials. Phys. Stat. Mech. Its Appl. 339, 145–151.

Sivanesapillai, R., Falkner, N., Hartmaier, A., Steeb, H., 2016. A CSF-SPH method for simulating drainage and imbibition at pore-scale resolution while tracking interfacial areas. Adv. Water Resour. 95, 212–234.

Sorbie, K.S., Skauge, A., 2012. Can network modeling predict two-phase flow functions? Petrophysics 53, 401–409.

Srisutthiyakorn, N., 2016. Deep-learning methods for predicting permeability from 2D/3D binary-segmented images. SEG Tech. Program Expand. Abstr. 2016 3042–3046.

Środoń, J., Drits, V.A., McCarty, D.K., Hsieh, J.C., Eberl, D.D., 2001. Quantitative X-ray diffraction analysis of clay-bearing rocks from random preparations. Clays Clay Miner. 49, 514–528.

Tartakovsky, A.M., Meakin, P., 2006. Pore scale modeling of immiscible and miscible fluid flows using smoothed particle hydrodynamics. Adv. Water Resour. 29, 1464–1478.

Valvatne, P.H., Blunt, M.J., 2004. Predictive pore-scale modeling of two-phase flow in mixed wet media. Water Resour. Res. 40.

Walderhaug, O., Eliassen, A., Aase, N.E., 2012. Prediction of permeability in quartz-rich sandstones: examples from the norwegian continental shelf and the fontainebleau sandstone. J. Sediment. Res. 82, 899–912.

Walls, J., Armbruster, M., 2012. Shale reservoir evaluation improved by dual energy X-Ray CT imaging. Technol. Update JPT Novemb.

Walls, J.D., Diaz, E., Cavanaugh, T., 2012. Shale reservoir properties from digital rock physics, in: SPE/EAGE European Unconventional Resources Conference & Exhibition-From Potential to Production.

Walls, J.D., Sinclair, S.W., 2011. Eagle Ford shale reservoir properties from digital rock physics. First Break 29, 97–101.

Wellington, S.L., Vinegar, H.J., 1987. X-ray computerized tomography. J. Pet. Technol. 39, 885–898.

Wildenschild, D., Roberts, J.J., Carlberg, E.D., 2000. On the relationship between microstructure and electrical and hydraulic properties of sand-clay mixtures. Geophys. Res. Lett. 27, 3085–3088.

Wildenschild, D., Sheppard, A.P., 2013. X-ray imaging and analysis techniques for quantifying pore-scale structure and processes in subsurface porous medium systems. Adv. Water Resour., 35th Year Anniversary Issue 51, 217–246. https://doi.org/10.1016/j.advwatres.2012.07.018

Yuan, H.H., Swanson, B.F., 1989. Resolving pore-space characteristics by rate-controlled porosimetry. SPE Form. Eval. 4, 17–24.

Zacharoudiou, I., Boek, E.S., 2016. Capillary filling and Haines jump dynamics using free energy Lattice Boltzmann simulations. Adv. Water Resour. 92, 43–56.